\documentclass[twocolumn,preprintnumbers,amsmath,amssymb,floatfix,superscriptaddress,aps,prc]{revtex4-1}
\usepackage{graphicx}% Include figure files
\usepackage{dcolumn}% Align table columns on decimal point
\usepackage{bm}% bold math
\usepackage{longtable}%two-column tables
\usepackage{tabularx}%advanced table formatting
\usepackage{lineno}

\usepackage{url}
\usepackage{hyperref}
    \hypersetup
    {
	pdfpagemode=None,
    colorlinks=true,
	linkcolor=blue,
	filecolor=blue,
	urlcolor=blue,
	citecolor=blue,
	bookmarks=false,
	pdftex=true,
	plainpages=false,
	hypertexnames=false,
	pdfpagelabels=true,
	hyperindex=true
    }

\begin{document}

\title{First-excited state $g$ factors in the stable, even Ge and Se isotopes}

\author{B.~P.~M$^{\rm c}$Cormick}
\affiliation{Department of Nuclear Physics, Research School of Physics, The Australian National University, Canberra, ACT 2601, Australia}
\author{A.~E.~Stuchbery}
\affiliation{Department of Nuclear Physics, Research School of Physics, The Australian National University, Canberra, ACT 2601, Australia}
\author{B.~A.~Brown}
\affiliation{Department of Physics and Astronomy, Michigan State University, East Lansing, MI 48824, USA}
\affiliation{National Superconducting Cyclotron Laboratory, Michigan State University, East Lansing, MI 48824, USA}
\author{G.~Georgiev}
\affiliation{Department of Nuclear Physics, Research School of Physics, The Australian National University, Canberra, ACT 2601, Australia}
\affiliation{CSNSM, CNRS/IN2P3, Universit\'e Paris-Sud, UMR8609, F-91405 ORSAY-Campus, France}
\author{B.~J.~Coombes}
\affiliation{Department of Nuclear Physics, Research School of Physics, The Australian National University, Canberra, ACT 2601, Australia}
\author{T.~J.~Gray}
\affiliation{Department of Nuclear Physics, Research School of Physics, The Australian National University, Canberra, ACT 2601, Australia}
\author{M.~S.~M.~Gerathy}
\affiliation{Department of Nuclear Physics, Research School of Physics, The Australian National University, Canberra, ACT 2601, Australia}
\author{G.~J.~Lane}
\affiliation{Department of Nuclear Physics, Research School of Physics, The Australian National University, Canberra, ACT 2601, Australia}
\author{T.~Kib\'{e}di}
\affiliation{Department of Nuclear Physics, Research School of Physics, The Australian National University, Canberra, ACT 2601, Australia}
\author{A.~J.~Mitchell}
\affiliation{Department of Nuclear Physics, Research School of Physics, The Australian National University, Canberra, ACT 2601, Australia}
\author{M.~W.~Reed}
\affiliation{Department of Nuclear Physics, Research School of Physics, The Australian National University, Canberra, ACT 2601, Australia}
\author{A.~Akber}
\affiliation{Department of Nuclear Physics, Research School of Physics, The Australian National University, Canberra, ACT 2601, Australia}
\author{L.~J.~Bignell}
\affiliation{Department of Nuclear Physics, Research School of Physics, The Australian National University, Canberra, ACT 2601, Australia}
\author{J.~T.~H.~Dowie}
\affiliation{Department of Nuclear Physics, Research School of Physics, The Australian National University, Canberra, ACT 2601, Australia}
\author{T.~K.~Eriksen}
\affiliation{Department of Nuclear Physics, Research School of Physics, The Australian National University, Canberra, ACT 2601, Australia}
\affiliation{Department of Physics, University of Oslo, Blindern, Oslo N-0316, Norway}
\author{S.~Hota}
\affiliation{Department of Nuclear Physics, Research School of Physics, The Australian National University, Canberra, ACT 2601, Australia}
\author{N.~Palalani}
\affiliation{Department of Nuclear Physics, Research School of Physics, The Australian National University, Canberra, ACT 2601, Australia}
\affiliation{Physics Department, University of Botswana, 4775 Notwane Rd., Gaborone, Botswana.}
\author{T.~Tornyi}
\affiliation{Department of Nuclear Physics, Research School of Physics, The Australian National University, Canberra, ACT 2601, Australia}

\date{\today}

\begin{abstract}
Transient-field $g$-factor measurements  in inverse kinematics were performed for the first-excited states of the stable, even isotopes of Ge and Se. The $g$~factors of $^{74}$Ge and $^{74}$Se were measured simultaneously using a cocktail beam, which eliminates most possible sources of systematic error in a relative $g$-factor measurement. The results are $g(^{74}{\rm Se})/g(^{74}{\rm Ge})=1.34(7)$,
$g(^{70}{\rm Ge})/g(^{74}{\rm Ge}) =  1.16(15)$,
$g(^{72}{\rm Ge})/g(^{74}{\rm Ge})=0.92(13)$,
$g(^{76}{\rm Ge})/g(^{74}{\rm Ge})=0.88(5)$,
$g(^{76}{\rm Se})/g(^{74}{\rm Se})=0.96(7)$,
$g(^{78}{\rm Se})/g(^{74}{\rm Se})=0.82(5)$,
$g(^{80}{\rm Se})/g(^{74}{\rm Se})=0.99(7)$ and
$g(^{82}{\rm Se})/g(^{74}{\rm Se})=1.19(6)$.
The measured $g$-factor ratios are in agreement with ratios from previous measurements, despite considerable variation in previous reported absolute values. The absolute values of the $g$~factors remain uncertain, however the Rutgers parametrization was used to set the transient-field strength and then compare the experimental $g$~factors with shell-model calculations based on the JUN45 and jj44b interactions. Modest agreement was found between experiment and theory for both interactions. The shell model calculations indicate that the $g(2^+_1)$ values and trends are determined largely by the balance of the spin carried by orbital motion of the protons.
\end{abstract}

\pacs{ }

\maketitle

\section{Introduction} \label{sect:Intro}

\subsection{Background and motivation}
The Ge and Se isotope chains are of interest as they span the $A \sim 70$ shape co-existence region \cite{Pe1990NPA,Pa2005PRL,Lj2008PRL,Ay2016PLB,Gu2019NPA}, exhibiting an interplay between single-particle and collective excitations, and shape changes \cite{Sp1998PRC,Pa2005PRL,Gu2013PRC,Gu2019NPA}. A microscopic understanding of the structure of these nuclei, through validated shell-model calculations, is desirable. These nuclei have recently become amenable to large-basis shell-model calculations \cite{Tsu2013JP,Shi2017PS} utilizing a $^{56}$Ni-core and the $p_{3/2}f_{5/2}p_{1/2}g_{9/2}$ basis space. Two effective interactions, JUN45 \cite{Ho2009PRC} and jj44b \cite{JJ44BUnpublished}, have been developed. Each uses different but overlapping data sets to define the interaction parameters. In order to test such calculations, accurate and precise $g$-factor systematics are needed. The $g(2^+_1)$ systematics in the even $^{70-76}$Ge and $^{74-82}$Se isotopes may be sensitive to sub-shell closures \cite{Be2007JPG,Me2003PRC}, providing a probe to test structural predictions as nuclear excitations shift from collective to single-particle behavior. However, the precise and accurate measurement of $g(2^+_1)$ in states such as these, with picosecond lifetimes, is challenging.

Along with this nuclear structure interest, the present work was motivated by the possibility to measure the $g$~factors of neutron-rich Ge isotopes by the recoil-in-vacuum (RIV) method.
Since the mid-2000s, there has been interest in the RIV technique \cite{RIVHIC} for measurements of nuclear $g$~factors for short-lived ($\lesssim$10$^{-7}$ s) states of Coulomb-excited radioactive-ion beams \cite{RIVTe2007}. For example, it has been used to measure the excited-state $g$~factors of several nuclides near doubly magic $^{132}$Sn \cite{132TeRIV,124SnRIV,134TeRIV,136TeRIV}. It has also given a precise $g$~factor for the $sd$-shell nucleus $^{24}$Mg \cite{Mg24RIV}, where very simple (H-like) atomic configurations are applicable. The present work developed from the intent to apply the RIV analysis method to Coulomb-excitation data taken on the neutron-rich isotopes $^{78-82}$Ge, like that obtained at the Holifield Radioactive Ion Beam Facility at Oak Ridge National Laboratory \cite{Pa2005PRL}. However, further experiments using the stable Ge and Se isotopes to calibrate the relevant free-ion hyperfine fields found an unexpected difference in the RIV behavior along the two isotope chains \cite{St2013HFI,RFLthesis}. Atomic structure effects may explain these differences, however the possibility also remains that the adopted $g$-factor ratios within and/or between the two isotope chains may be responsible for some, or all, of the observed difference. The desire to verify previous measurements of the relative $g$~factors along and between the Ge and Se isotope chains motivated the present work, for both the nuclear structure interest and to aid the interpretation of the RIV data. To this end, the relative $g$~factors of the stable, even isotopes of Ge and Se were measured, and the first case of a relative $g$-factor measurement in inverse kinematics using a cocktail beam of $^{74}$Ge and $^{74}$Se was performed.

\subsection{Transient-field $g$-factor measurements}
When an ion with a velocity of a few percent of the speed of light traverses a polarized ferromagnetic foil, it experiences a strong transient hyperfine field on the order of a few kilotesla. This transient field (TF) has been utilized to perform $g$-factor measurements on nuclear states with picosecond lifetimes since the mid-1970s \cite{TFAnnRev,Be2007JPG}. While the magnitude of the transient field makes these measurements possible, a quantitative microscopic understanding of the field's origin and strength has still not been achieved \cite{Sp2002PPNP,Be2007JPG}. As such, determination of the absolute value of a $g$~factor often relies on parametrizations, which have inherent uncertainties. They are generally considered to be accurate to about $\pm 10\%$, but larger deviations can occur \cite{Ch2011PRC}. Therefore, experiments utilizing the TF technique are most reliable when measuring $g$~factors relative to a nucleus which is nearby on the nuclear chart and has an already known, independently determined $g$~factor. The difficulty for the Ge and Se isotopes, with $Z=32$ and 34, respectively, is that there is no nearby suitable calibration state with a sufficiently well known $g$~factor. These isotopes fall between two relatively precise calibration points used to establish the empirical parametrizations \cite{RutgersTFParam,EbTFParam}, namely the first-excited states of $^{56}$Fe ($Z=26$) and $^{106}$Pd ($Z=46$). (See Refs.~\cite{Ea2009PRCFe56,Ch2011PRC} for further discussion.)

Although the absolute values of $g$~factors determined by the TF method can be uncertain, relative values can be reliably determined. In particular, systematic uncertainties in relative $g$~factors can be reduced by performing measurements using the same target foil, and most ideally when two or more isotopes are measured simultaneously so that any changes in the target, such as those induced by beam heating, affect both measurements equally. Relative $g$-factor measurements are the primary objective in the present work.

The $g$~factors of the first-excited states in the stable, even Ge isotopes have been measured repeatedly by different groups over the years using the ion-implantation perturbed-angular-correlation (IMPAC) \cite{He1969NPA} and TF  \cite{Pa1984JPG,La1987AJP,Gu2013PRC} techniques. The most recent measurement performed on the Ge isotopes, by G\"{u}rdal \textit{et al.} \cite{Gu2013PRC}, provided the most precise values yet, with $g(2^+_1)$ determined using the TF technique in inverse kinematics and the Rutgers parametrization \cite{RutgersTFParam}. An older set of measurements, performed by Pakou \textit{et al.} \cite{Pa1984JPG}, had comparable precision, and used the same parametrization. However, Pakou \textit{et al.} used iron hosts and conventional kinematics (target excitation), whereas G\"{u}rdal \textit{et al.} used gadolinium hosts and inverse kinematics (beam excitation). The two approaches deduced significantly different $g(2^+_1)$ values in $^{74,76}$Ge. This difference raises the question as to whether the disagreement arises from the experimental method, or the accuracy of the parametrization, or both.

The number of independent $g$-factor measurements of the first-excited states in the stable, even Se isotopes is much fewer than for Ge, with the most recent measurement performed by Speidel \textit{et al.} in 1998 \cite{Sp1998PRC}. Absolute $g(2^+_1)$ values were determined from their relative $g$-factor measurements by adopting $g(2^+_1;\mathrm{^{82}Se})=+0.496(29)$ as the reference. This absolute $g$~factor was deduced from a previous measurement of $^{82}$Se $2^+_1$-state precession angles in iron and gadolinium hosts \cite{Br1978HI}. As the precession angles from Ref.~\cite{Br1978HI} were measured to study the TF, not to determine a $g$~factor, it appears that the adopted $g$~factor was obtained by re-analyzing those data using the Rutgers TF parametrization to calibrate the field strength. As such, this calibration $g$~factor is not an independently determined quantity, and the accuracy of the magnitude of the reported $g$~factors due to the normalizing $g$~factor remains in question.

\subsection{Present work}
In the present work, a series of relative TF measurements of first-excited state $g$~factors in the stable, even Ge and Se isotopes is reported. A rigorous relative measurement between $g(2^+_1;\mathrm{^{74}Se})$ and $g(2^+_1;\mathrm{^{74}Ge})$ was performed by simultaneous TF measurements using cocktail beams. The other stable, even isotopes of Ge and Se were measured using the same targets and during the same experimental runs as the $A=74$ isotopes. The measured $g$-factor ratios are compared to those from previously published measurements, and precise relative values for the $g$~factors are derived from a weighted average of the present and previous measurements. Absolute $g$~factors were determined from measured precession angles and the Rutgers parametrization \cite{RutgersTFParam}, and are compared to large-basis shell-model calculations.

The paper is arranged as follows. Experimental procedures are described in Sect.~\ref{sect:Methods}. Experimental results are described in Sect.~\ref{sect:Results}. A discussion of these results in light of previous work and shell-model calculations is presented in Sect.~\ref{sect:Discussion}. The results are summarized and conclusions drawn in Sect.~\ref{sect:Conclusion}.

\section{Experimental Procedures} \label{sect:Methods}

\begin{table*}[t]
\centering
\caption{Summary of measurements.}
\label{tab:runs}
\begin{tabularx}{\hsize}{rccccc}
\hline
\hline
 & Run 1 & Run 2\footnotemark[1] & Run 3 & Run 4\footnotemark[2] & Run 5\footnotemark[2] \\
\hline
Beam Energy (MeV) & 190 & 180 & 190 & 190 & 190 \\
Beam Species & $^{70,72,74}$Ge, $^{76,78,80,82}$Se & $^{74,76}$Ge & $^{74}$Ge + $^{74}$Se\footnotemark[3] & $^{80,82}$Se & $^{78,82}$Se \\
 & $^{74}$Ge + $^{74}$Se\footnotemark[3] & & & & \\
Target (C/Fe[Gd]/Cu) & 0.48/3.66/5.6 & 0.37/3.67/10.6 & 0.28/4.23/3.94 & 0.28/4.23/3.9 & 0.5/[5.92]/5.5 \\
(mg/cm$^2$) & & & & & \\
\\
$\phi_p$  & $90^{\circ}$ & $90^{\circ}$ & $0^{\circ}, 45^{\circ}, 90^{\circ},$ & $0^{\circ}, 45^{\circ}, 90^{\circ},$ & $0^{\circ}, 45^{\circ}, 90^{\circ},$ \\
 & & & $135^{\circ}, 180^{\circ}$ & $135^{\circ}, 180^{\circ}$ & $135^{\circ}, 180^{\circ}$  \\
\\
$\theta_{\gamma}$,  precession & $\pm 65^{\circ}, \pm 115^{\circ}$ & $\pm 65^{\circ}, \pm 115^{\circ}$ & $\pm 50^{\circ}, \pm 130^{\circ}$ & $\pm 65^{\circ}, \pm 115^{\circ}$ & $\pm 45^{\circ}, \pm 135^{\circ}$ \\
\\
$\theta_{\gamma}$, angular correlation & $0^{\circ}, \pm 30^{\circ}, \pm 45^{\circ},$ & $0^{\circ}, \pm 30^{\circ}, \pm 40^{\circ},$ & $0^{\circ}, \pm 30^{\circ}, \pm 45^{\circ},$ & $0^{\circ}, \pm 25^{\circ}, \pm 55^{\circ}$ & \\
 & $ \pm 55^{\circ}, \pm 60^{\circ}, \pm 65^{\circ}$ & $\pm 45^{\circ}, \pm 50^{\circ}, \pm 55^{\circ},$ & $ \pm 50^{\circ}, \pm 55^{\circ}, \pm 65^{\circ},$ & $\pm 65^{\circ}, \pm 88^{\circ}$ & \\
 & & $\pm 60^{\circ}, \pm 65^{\circ}, \pm 75^{\circ},$ & $\pm 75^{\circ}$ & & \\
 & & $\pm 115^{\circ}, 235^{\circ}, \pm 135^{\circ}$ & & & \\
\hline
\hline
\end{tabularx}
\begin{flushleft}
\footnotetext[1]{ Runs 1 and 2 were separated in time because the ion source was contaminated after running the Se and cocktail beams, and the $\gamma$-ray energies associated with $^{76}$Ge and $^{76}$Se, 562.9 and 559.1 keV, respectively, are too close to separate reliably.}
\footnotetext[2]{ Runs 4 and 5 were consecutive, beginning with the $^{80}$Se beam, followed by $^{82}$Se and $^{78}$Se beams. The target was changed during delivery of the $^{82}$Se beam.}

\footnotetext[3]{ Cocktail beam.}
\end{flushleft}
\end{table*}

\begin{table*}[t]
\centering
\caption{Measured transient-field precession angles and reaction kinematics for the $2^+_1$ states of Ge and Se isotopes as their ions traverse the ferromagnetic layer of the target. $E_i$($E_e$) is average energy at entry into (exit from) the foil, $v_i$($v_e$) is the average velocity at entry into (exit from) the foil, $\langle v\rangle$ is the average velocity through the foil, $T$ is the effective transit time, and $v_0 = c/137$ is the Bohr velocity. $\tau$ is the mean life of the 2$^+_1$ state. $\Phi(\tau)$ is evaluated from Eq.~(\ref{eq:phi}) with the velocity-dependent transient-field strength given by Eq.~(\ref{eq:Btf}) using the Rutgers parameters \cite{RutgersTFParam} (see text). Fig.~\ref{fig:tripledetector} shows the definition of $\langle \theta _p \rangle$ for runs~1~and~2.}

\label{tab:kinematics}
\begin{tabularx}{\textwidth}{lXXXXXXXcc}
\hline
\hline
Nuclide & $E_i$ (MeV) & $E_e$ (MeV) & $v_i$/$v_0$ & $v_e$/$v_0$ & $\langle v$/$v_0\rangle$ & $T$(fs) & $\tau$(ps)\footnotemark[1] & $-\Phi(\tau)$ (mrad) & $-\Delta\theta$ (mrad)  \\
%\cline{9-11}
\hline
Run 1:  \\
$\langle\theta_{\rm p}\rangle=22^{\circ}$  \\
$^{70}$Ge & 91.2 & 10.9 & 7.25 & 2.50 & 4.61 & 407 & 1.91 & 29.3  & 8.4(13) \\
$^{72}$Ge & 92.6 & 11.9 & 7.20 & 2.58 & 4.59 & 444 & 4.75 & 31.8  & 7.3(12) \\
$^{74}$Ge & 93.9 & 12.9 & 7.15 & 2.65 & 4.59 & 462 & 18.1 & 33.2  & 9.2(8) \\
$^{74}$Ge & 93.9 & 12.9 & 7.15 & 2.65 & 4.59 & 462 & 18.1 & 33.2  & 9.5(12) \footnotemark[2] \\
$^{74}$Se & 92.8 & 10.1 & 7.11 & 2.35 & 4.37 & 480 & 10.2 & 35.9  & 13.3(9)\footnotemark[2] \\
$^{76}$Se & 94.1 & 11.1 & 7.07 & 2.43 & 4.41 & 481 & 17.8 & 36.2  & 12.4(8) \\
$^{78}$Se & 95.5 & 12.1 & 7.03 & 2.50 & 4.45 & 474 & 14.0 & 35.9  & 10.1(7) \\
$^{80}$Se & 96.9 & 13.1 & 6.99 & 2.57 & 4.49 & 467 & 12.3 & 35.6  & 13.3(9) \\
$^{82}$Se & 98.1 & 14.1 & 6.94 & 2.63 & 4.51 & 468 & 18.6 & 35.8  & 15.2(10)\\
Run 1: \\
$\langle\theta_{\rm p}\rangle=0^{\circ}$  \\
$^{70}$Ge & 80.2 & 7.16 & 6.79 & 2.03 & 4.06 & 452 & 1.91 & 30.7  & 11.1(16) \\
$^{72}$Ge & 81.8 & 8.04 & 6.77 & 2.12 & 4.06 & 495 & 4.75 & 33.5  & 9.4(16) \\
$^{74}$Ge & 83.4 & 8.95 & 6.74 & 2.21 & 4.09 & 513 & 18.1 & 35.0  & 7.6(13) \\
$^{74}$Ge & 83.4 & 8.95 & 6.74 & 2.21 & 4.09 & 513 & 18.1 & 35.0  & 8.4(17) \footnotemark[2] \\
$^{74}$Se & 82.3 & 6.78 & 6.69 & 1.92 & 3.86 & 537 & 10.2 & 38.0  & 13.7(12)\footnotemark[2] \\
$^{76}$Se & 83.9 & 7.61 & 6.67 & 2.01 & 3.92 & 536 & 17.8 & 38.2  & 14.2(10) \\
$^{78}$Se & 85.4 & 8.49 & 6.64 & 2.09 & 3.99 & 523 & 14.0 & 37.7  & 12.9(9) \\
$^{80}$Se & 87.0 & 9.39 & 6.62 & 2.17 & 4.04 & 514 & 12.3 & 37.3  & 13.6(13) \\
$^{82}$Se & 88.4 & 10.3 & 6.59 & 2.25 & 4.08 & 513 & 18.6 & 37.4  & 16.3(13)\\
\\
Run 2:  \\
$\langle\theta_{\rm p}\rangle=22^{\circ}$  \\
$^{74}$Ge & 91.0 & 11.5 & 7.04 & 2.51 & 4.44 & 478 & 18.1 &  33.8 & 9.2(5) \\
$^{76}$Ge & 92.3 & 12.5 & 7.00 & 2.58 & 4.48 & 477 & 26.9 &  33.9 & 7.9(5) \\
Run 2:   \\
$\langle\theta_{\rm p}\rangle=0^{\circ}$   \\
$^{74}$Ge & 82.0 & 8.39 & 6.68 & 2.14 & 4.01 & 525 & 18.1 &  35.4 & 10.7(7) \\
$^{76}$Ge & 83.6 & 9.29 & 6.66 & 2.22 & 4.07 & 520 & 26.9 &  35.4 & 9.7(6) \\
\\
Run 3: \\
$^{74}$Ge & 111 & 13.0 & 7.76 & 2.66 & 4.87 & 505 & 18.1 &  37.2  & 9.5(5)\footnotemark[3] \\
$^{74}$Se & 110 & 10.1 & 7.74 & 2.34 & 4.63 & 525 & 10.2 &  40.3  & 13.5(6)\footnotemark[3]  \\
\\
Run 4:   \\
$^{80}$Se & 114 & 13.0 & 7.57 & 2.56 & 4.72 & 516 & 12.3 &  40.1  & 13.8(10) \\
$^{82}$Se & 115 & 13.9 & 7.51 & 2.62 & 4.74 & 517 & 18.6 &  40.3  & 17.3(15) \\
\\
Run 5:   \\
$^{78}$Se & 103 & 18.8 & 7.30 & 3.12 & 4.96 & 680 & 14.0 &  54.3  &  17.6(8) \\
$^{82}$Se & 106 & 21.1 & 7.20 & 3.22 & 4.98 & 681 & 18.6 &  54.5  & 25.9(13) \\
\hline
\hline
\end{tabularx}
\begin{flushleft}
\footnotetext[1]{From Ref.~\cite{RamanBE2}. Uncertainties in $\tau$ can be neglected in the evaluation of $\Phi(\tau)$.}
\footnotetext[2]{Simultaneous measurement of $^{74}$Ge and $^{74}$Se with cocktail beam.}
\footnotetext[3]{Simultaneous measurement of $^{74}$Ge and $^{74}$Se with cocktail beam.}
\end{flushleft}
\end{table*}

\begin{figure}[]
\centerline{
  \includegraphics[width=\columnwidth]{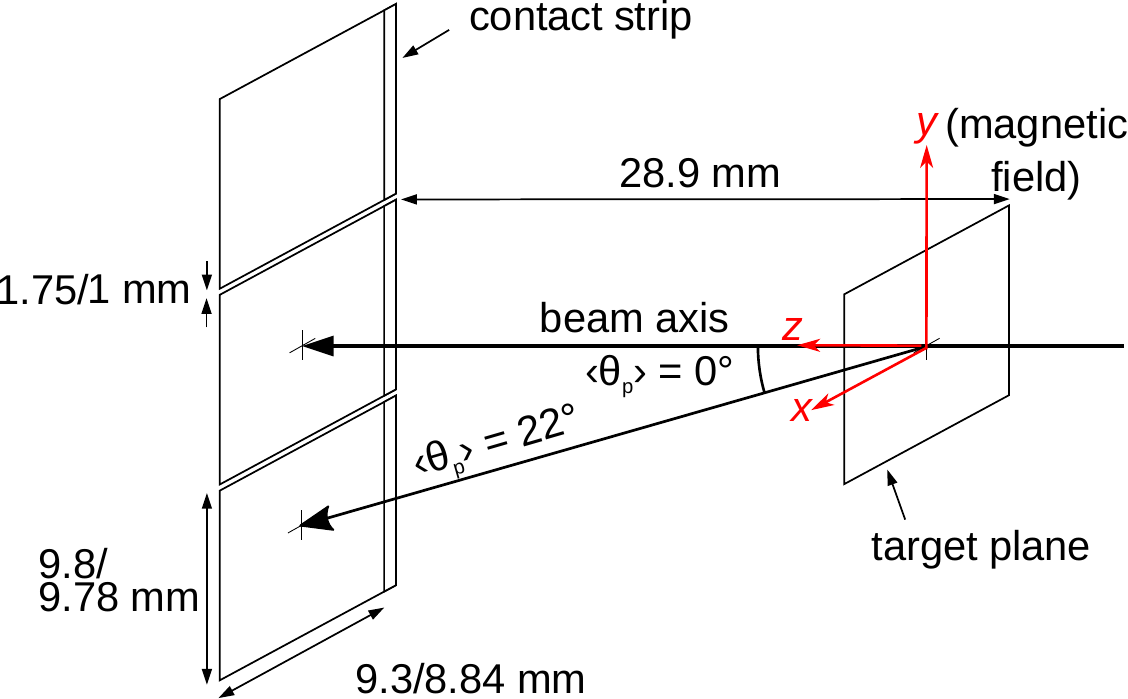}
}
\caption{Experimental geometry for run~1 and run~2 showing the beam axis and the particle detector dimensions (run~1~/~run~2). The average scattering angles for the $^{12}$C target ions are indicated by $\langle \theta_{p} \rangle$.}
\label{fig:tripledetector}
\end{figure}

\begin{figure}[]
\centerline{
  \includegraphics[width=\columnwidth]{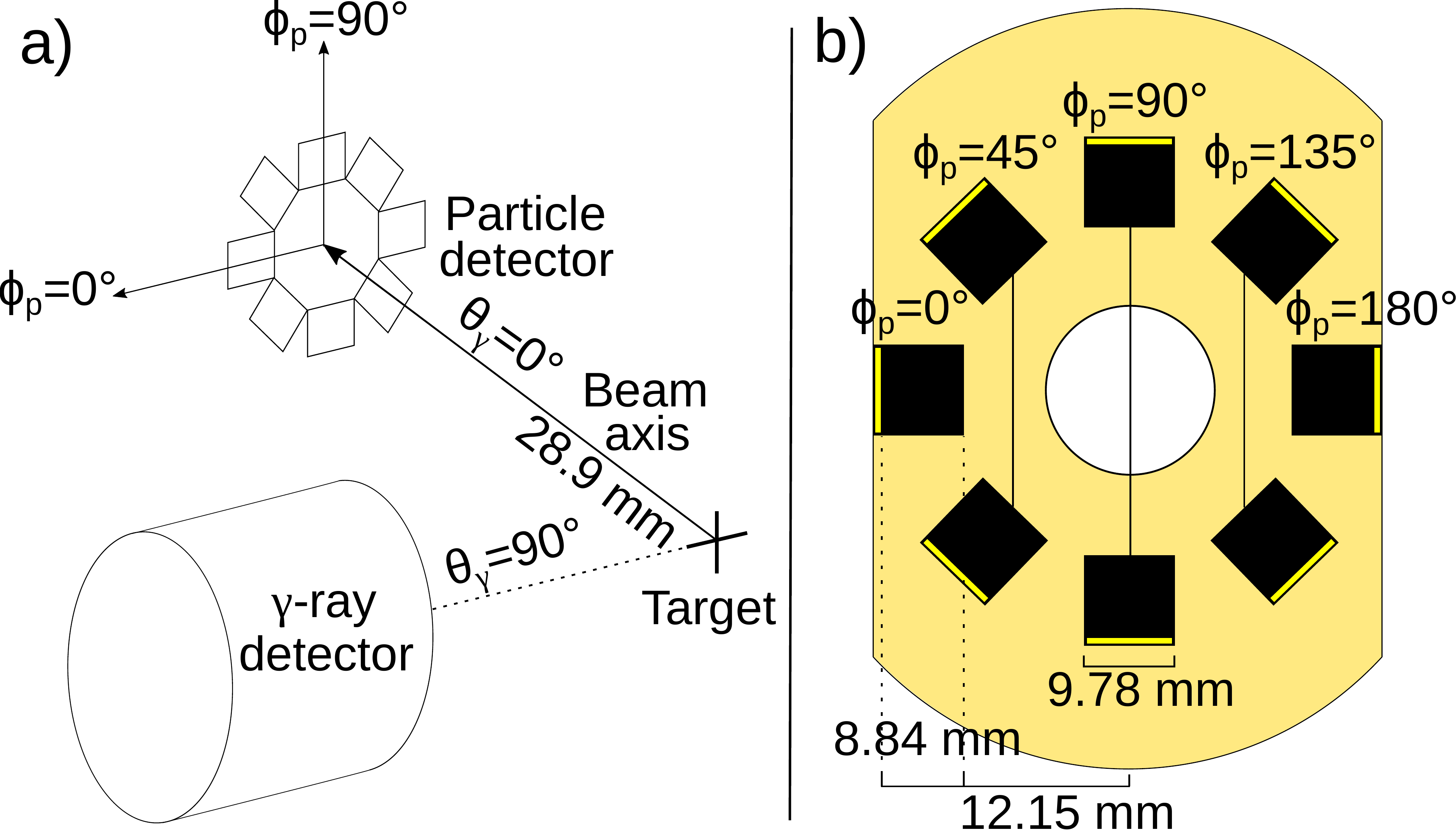}
}
\caption{a) Experimental geometry for runs $3-5$ showing the beam axis, a $\gamma$-ray detector and the particle detectors. b) A schematic of the particle detector array indicating dimensions and joined detectors.}
\label{fig:heliotrope}
\end{figure}

\begin{figure}[]
\centerline{
  \includegraphics[width=\columnwidth]{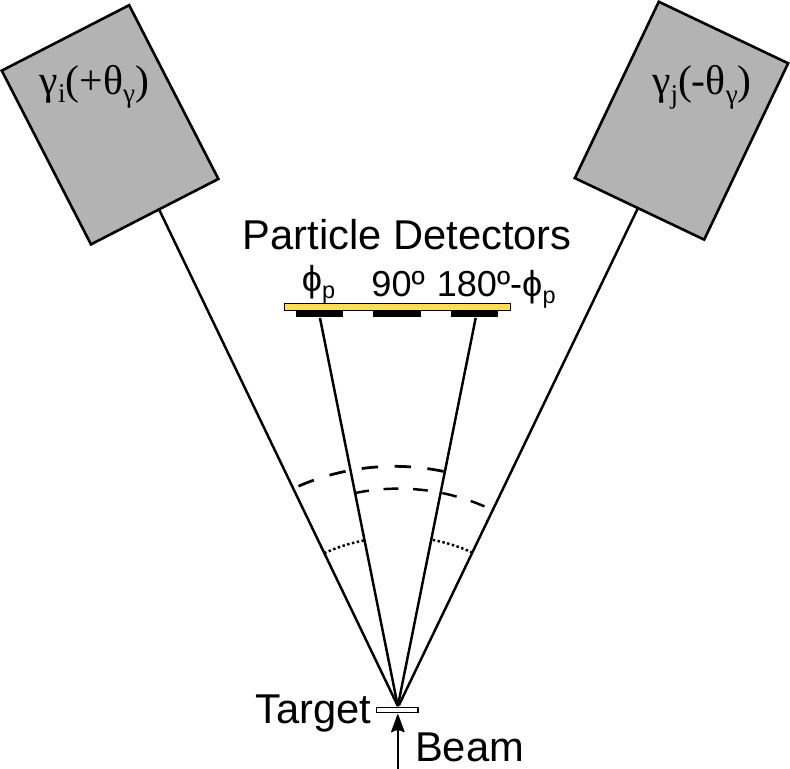}
}
\caption{Illustration of the particle-$\gamma$ coincidence geometry, drawn from a top-down view. Symmetry is maintained when particle-$\gamma$ coincidence pairs are formed with $\gamma_i$ at $+\theta$ paired with the particle detector at $\phi_p$ and $\gamma_j$ at $-\theta$ paired with the particle detector at $180^{\circ}-\phi_p$, shown by the equivalent dotted vs dashed angles. In short, $\gamma_i(\phi_p)$ must be paired with $\gamma_j (180^{\circ}-\phi_p)$. Fig.~\ref{fig:heliotrope} shows the definition of the angle $\phi_p$.}
\label{fig:heliosym}
\end{figure}

The transient-field method in inverse kinematics was employed to measure relative $g(2^+_1)$ values in the stable, even isotopes of Ge and Se. The technique has been reviewed in Refs.~\cite{Sp2002PPNP,Be2007JPG}, and previous measurements by this method using the Australian National University (ANU) Hyperfine Spectrometer \cite{ANUHIAF} have been described in detail elsewhere \cite{Ea2009PRCFe56,Ea2009PRCFerel,Ch2011PRC,Mc2018PLB}. Data were collected over five separate runs, the details of which are listed in Table~\ref{tab:runs}. Ion beams were delivered by the 14UD Pelletron accelerator at the ANU Heavy Ion Accelerator Facility. The beams were incident upon a multi-layer target consisting of a front $^{\rm nat}$C layer, an iron or gadolinium foil, and a copper backing. The target details are presented in Table~\ref{tab:runs}. The front $^{\rm nat}$C layer served to Coulomb-excite the beam, the central iron or gadolinium foil served as the ferromagnetic host for the TF precession effect, and the copper backing served as a ``field-free'' environment to stop the recoiling beam particles. Relevant details of the reaction kinematics are summarized in Table~\ref{tab:kinematics}. A cryocooler kept the target at $\sim$5~K. An external magnetic field of $\sim$0.09~T was applied in the vertical direction to polarize the ferromagnetic layer of the target, and its direction was reversed periodically ($\sim$15~min).

%In the case of $^{76}$Ge, a separate run was required because

\begin{figure}[]
\centerline{
  \includegraphics[width=\columnwidth]{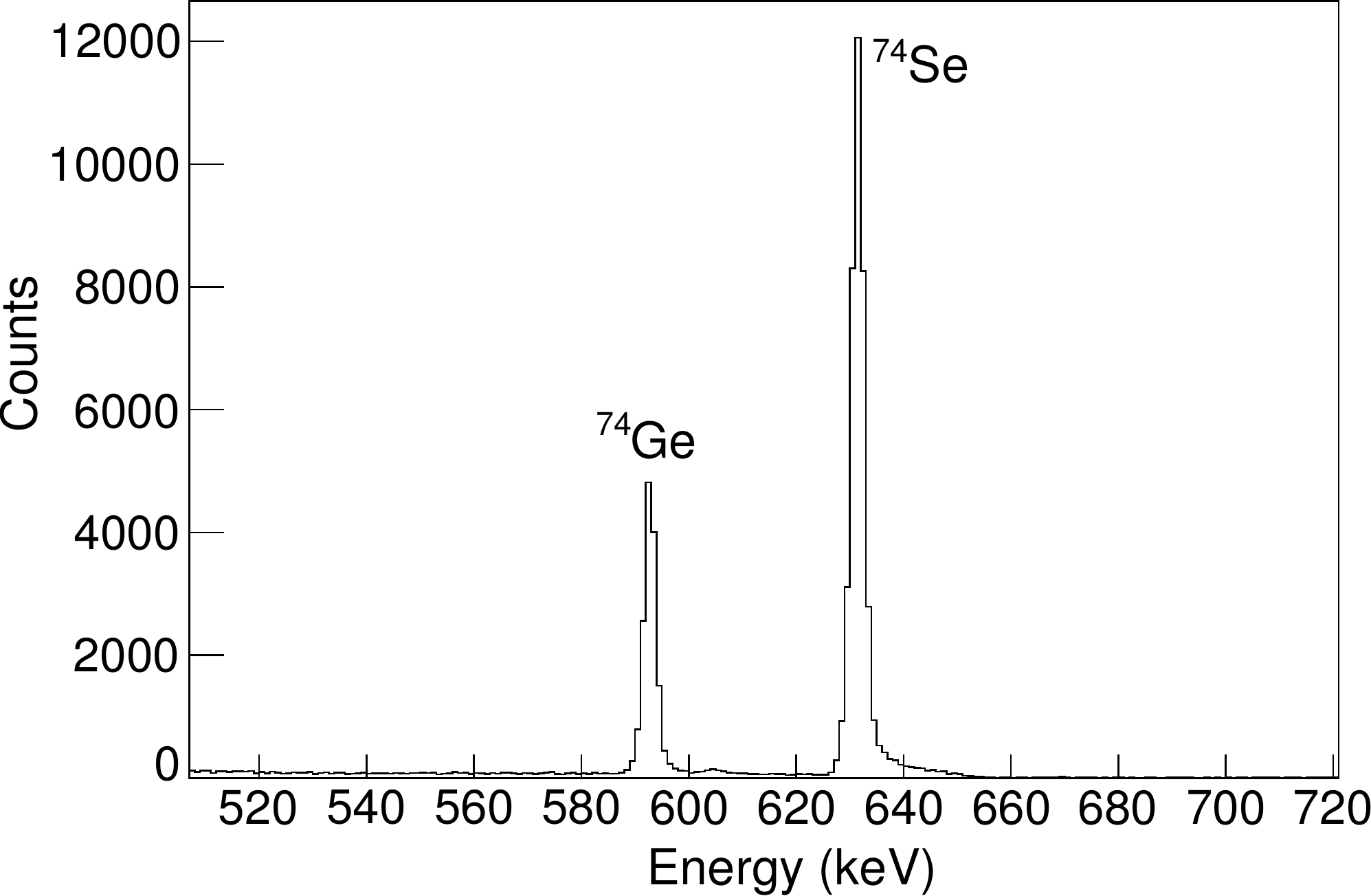}
}
\caption{Spectrum of $\gamma$ rays at $65^{\circ}$ to the beam in coincidence with $^{12}$C recoils measured during run~1 with the $^{74}$Ge/$^{74}$Se cocktail beam. Random coincidences have been subtracted. The resulting spectrum has almost no background. This spectrum represents $\sim$25\% of the data taken for the simultaneous measurement of $^{74}$Ge and $^{74}$Se during run~1.}
\label{fig:photopeak}
\end{figure}

Forward-scattered C ions were detected by silicon photodiodes, arranged as shown in Fig.~\ref{fig:tripledetector} for runs~1~and~2, and Fig.~\ref{fig:heliotrope} for runs $3-5$. The central detector for runs~1 and~2 was covered with a 1.7-mg/cm$^2$-thick mylar sheet to stop any transmitted beam, knock-on target ions, or low-energy electrons ($\delta$~rays). Four HPGe detectors, positioned in the horizontal plane, detected $\gamma$~rays emitted from the Coulomb-excited beam particles.
Average particle-detector azimuthal angles ($\phi_p$) and $\gamma$-ray detector angles ($\theta_{\gamma}$) for each run are summarized in Table~\ref{tab:runs}. The co-ordinate frame defining these particle- and $\gamma$-detector angles has its origin at the beam spot on the target. The beam direction defines the $z$ or polar axis ($\theta = 0$) and the magnetic field is applied along the $y$ axis, which is along $(\theta,\phi) = (90^{\circ},90^{\circ})$  %$(\theta,\phi) = (\pi/2 ,\pi/2)$
in spherical polar coordinates. The $\gamma$-ray detectors are located in the $xz$ or $\phi=0$ plane, and the particle detectors are positioned parallel to the $xy$ plane. In the following discussion of runs 3-5 the particle detectors are identified by their $\phi_p$ angle as indicated in Fig.~\ref{fig:heliotrope}b.

The $\gamma$-ray~detector angles employed for the TF precession measurements were selected to maximize the measured precession effect for the available particle-detector angles \cite{NIM-paper}. Angular correlations were measured by moving two or three $\gamma$-ray detectors through a sequence of different angles with respect to the beam, while the remaining detector(s) remained fixed. The angular correlation data sets were normalized using particle-$\gamma$ coincidence counts in the fixed $\gamma$-ray~detector(s). The data from runs~1~and~2 were recorded using a recently acquired XIA Pixie-16 digital pulse processor \cite{XIAPixie16}; these results are the first published from this installation.

Angular precessions were measured using standard procedures \cite{Be2007JPG}, with some modification for the annular arrangement of the particle detectors in runs 3-5. The TF induces a rotation, $\Delta\theta$, in the angular correlation of the de-excitation $\gamma$ rays, $W(\theta)$, for the nuclei traversing the ferromagnetic medium. Double ratios of observed counts were formed:
\begin{equation}\label{eq:rhoij}
\rho_{ij} = \sqrt{\frac{N(\theta_i)\uparrow}{N(\theta_i)\downarrow}\frac{N(\theta_j)\downarrow}{N(\theta_j)\uparrow}},
\end{equation}
where $N(\theta_i)$ and $N(\theta_j)$ represent particle-$\gamma$ coincidence counts in the photopeak measured in $\gamma$-ray detectors $i$ and $j$ at angles $+\theta_{\gamma}$ and $-\theta_{\gamma}$, respectively, and $\uparrow$ and $\downarrow$ represent the field direction.

For runs~1 and~2 the detectors had the usual particle-$\gamma$ coincidence symmetry with the particle detectors at $\phi_p=90^{\circ}$ and $\phi_p=270^{\circ}$ (sometimes designated $\phi_p=-90^{\circ}$), so double ratios can be formed in the conventional way \cite{TFAnnRev} for each of the three particle detectors.

% $\phi=\pi/2$ (and $\phi=3\pi/2$)

To calculate $\rho_{ij}$ for runs 3-5 using the detector type shown in Fig.~\ref{fig:heliotrope}, a modified procedure is required when using counts from the photodiodes with $\phi_p \neq \pm 90^{\circ}$. The symmetries required to form the particle-$\gamma$ coincidences in Eq.~(\ref{eq:rhoij}) can be inferred from Fig.~\ref{fig:heliosym}, which represents a top-down view of the particle and $\gamma$-ray detectors projected onto the horizontal plane through the $\gamma$-ray detectors. For clarity, Fig.~\ref{fig:heliosym} shows only one pair of particle detectors at angles $\phi_p$ and $180^{\circ}-\phi_p$. By inspection of Fig.~\ref{fig:heliosym} it can be seen that the relative angle between the $\gamma$-ray detector at $+\theta_{\gamma}$ and the particle detector at $\phi_p$ is equivalent to that between the $\gamma$-ray detector at $-\theta_{\gamma}$ and the particle detector at $180^{\circ}-\phi_p$, as indicated by the dotted arcs. There is another pair of particle-$\gamma$ combinations with the relative angle indicated by the dashed arcs in Fig.~\ref{fig:heliosym}; the angle between detectors at $+\theta_{\gamma}$ and $180^{\circ}-\phi_p$  is equivalent to that between the detectors at $-\theta_{\gamma}$ and  $\phi_p$.

These symmetries indicate that Eq.~(\ref{eq:rhoij}) should be constructed such that if $N(\theta_i)$ represents $N(\theta_i,\phi_p)$, then $N(\theta_j)$ must represent $N(\theta_j,180^{\circ} - \phi_p)$, and if $N(\theta_i)$ represents $N(\theta_i,180^{\circ}-\phi_p)$, then $N(\theta_j)$ must represent $N(\theta_j,\phi_p)$. A more detailed account of this procedure will be published elsewhere \cite{NIM-paper}.

The spin-rotation (precession) angle $\Delta\theta$ is determined from:
\begin{equation}\label{eq:dthexp}
\Delta\theta = \frac{\epsilon}{S},
\end{equation}
where
\begin{equation}\label{eq:eps}
\epsilon = \frac{1-\rho}{1+\rho},
\end{equation}
and $S$ is the logarithmic derivative of the angular correlation at $+\theta_{\gamma}$
\begin{equation}\label{eq:slope}
S = S(\theta_{\gamma}) = \left . \frac{1}{W}\frac{dW}{d\theta} \right |_{\theta_\gamma}.
\end{equation}

\begin{figure}[]
\centerline{
  \includegraphics[width=\columnwidth]{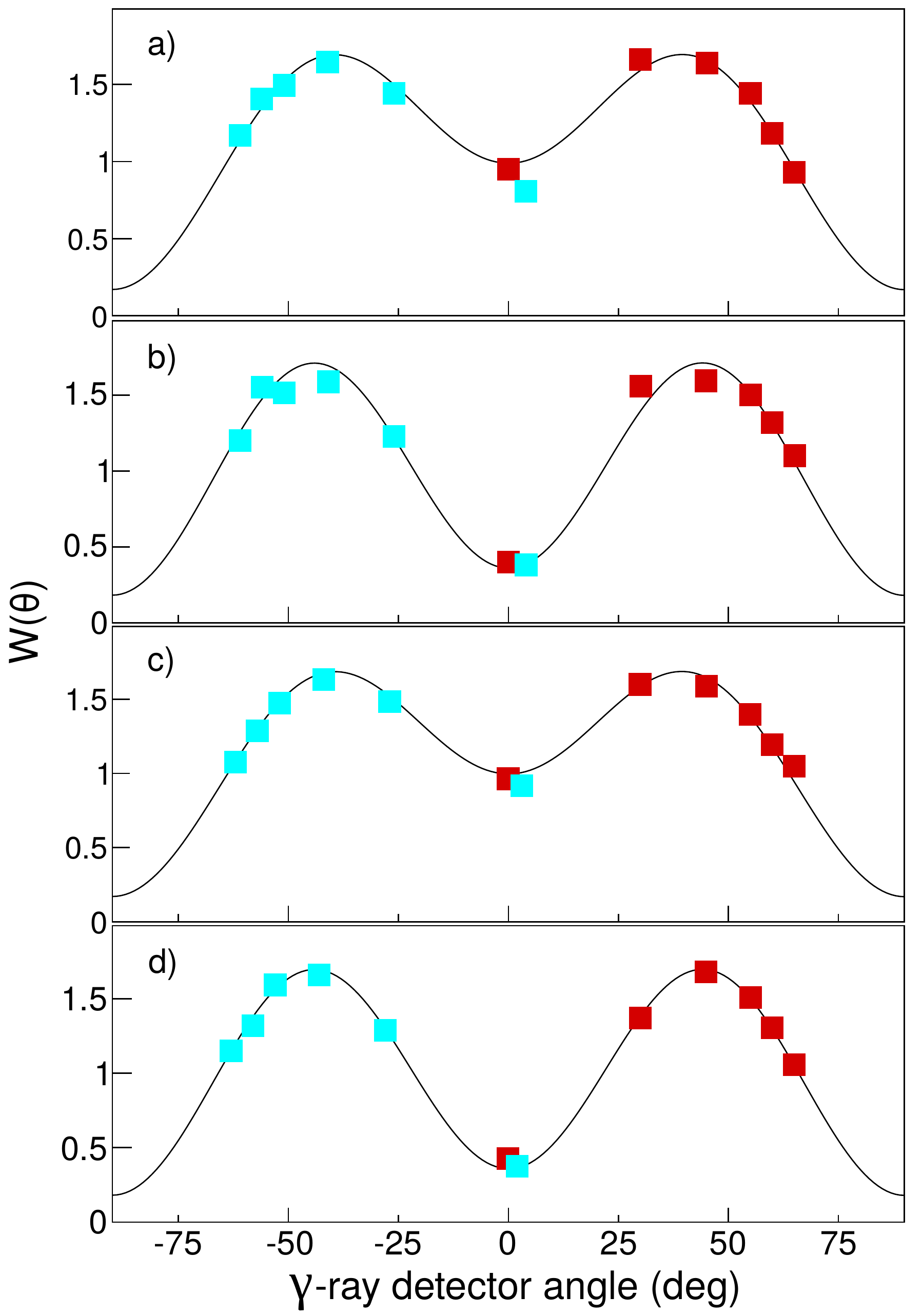}}
\caption{Angular correlations for the $^{74}$Ge and $^{74}$Se $2^+_1 \rightarrow 0^+_1$ transitions from run~1 for the two unique particle-detector angles. a) $^{74}$Ge, top and bottom particle detectors. b) $^{74}$Ge, center particle detector. c) $^{74}$Se, top and bottom particle detectors. d) $^{74}$Se, center particle detector. Statistical errors are smaller than or similar in size to the data points. Dark red (light blue) points correspond to the $\gamma$-ray detector that moves to positive (negative) angles. The measured data are normalized to the theoretical angular correlations shown as continuous lines. Although there is a distinct difference between the correlations for the top/bottom and center particle detectors near 0$^{\circ}$, they are similar near $\pm$65$^{\circ}$, meaning the two sets of particle detectors have similar sensitivity for the $g$-factor measurement. An offset of +3$^{\circ}$ was required on the negative angle data (all taken with the same $\gamma$-ray detector) to optimize the fit.
}
\label{fig:GeSe18AC}
\end{figure}

\begin{figure}[]
\centerline{
  \includegraphics[width=\columnwidth]{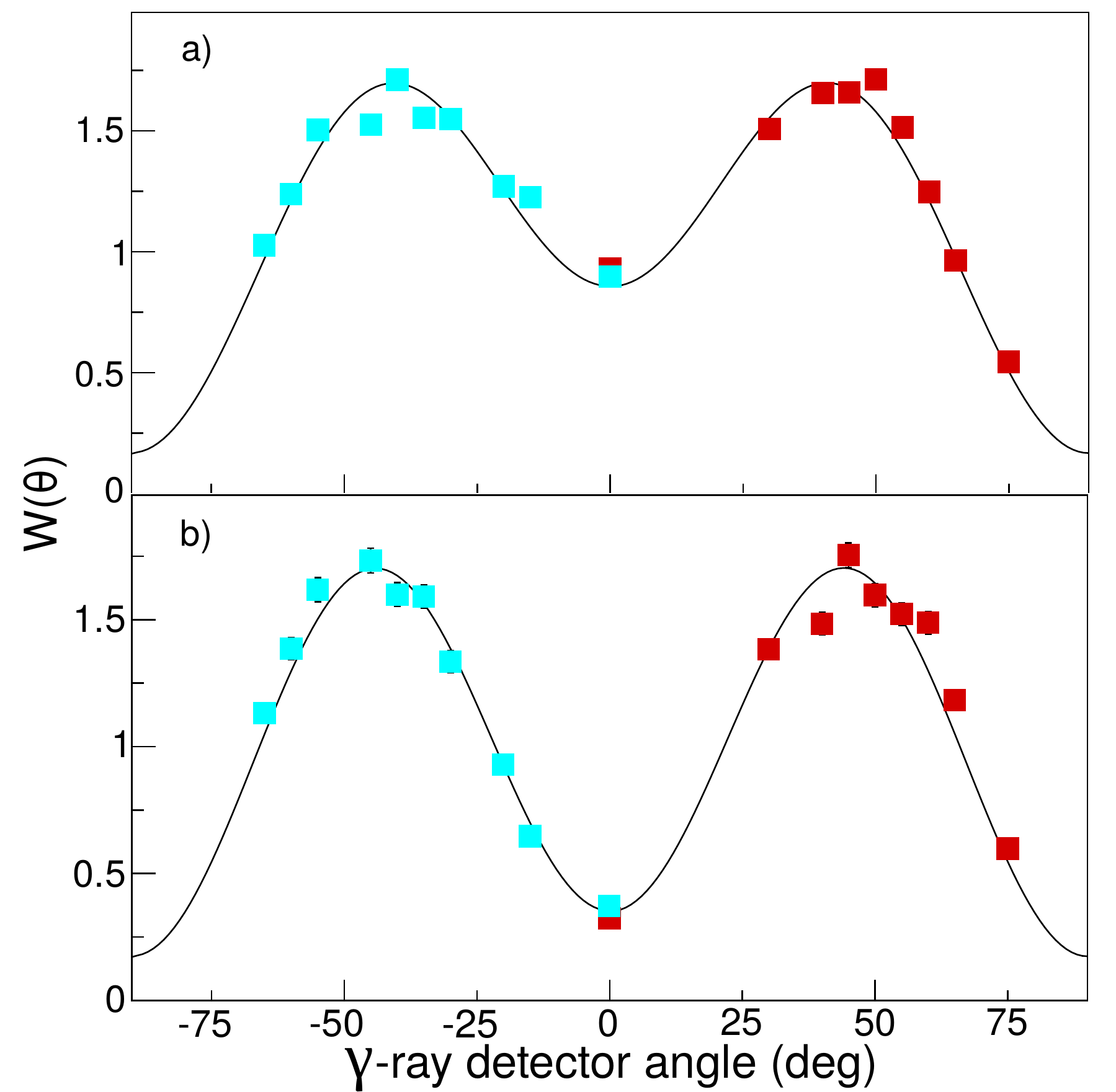}
}
\caption{Angular correlations for the $^{76}$Ge $2^+_1 \rightarrow 0^+_1$ transition from run~2 for the two unique particle-detector angles.  a) Top and bottom particle detectors. b) Center particle detector. See also the caption of Fig.~\ref{fig:GeSe18AC}.
 }
\label{fig:Ge76AC}
\end{figure}

Along with the double ratios used to determine the TF precession effect it is useful to form ``cross ratios", which should show no net effect, as a check on the validity of the data. For the conventional ($\phi_p=90^{\circ}$) particle detector geometry cross ratios are given by Eq.~(\ref{eq:rhoij}) with $\theta_i=\theta_{\gamma}$ and $\theta_j=180^{\circ}-\theta_{\gamma}$, for example. Such symmetrically placed particle and $\gamma$-ray detectors should have $\rho$ values distributed around unity (or $\epsilon$ distributed around zero); departures from the null effect can indicate systematic errors.

The angular correlations were computed using routines based on the Winther-de~Boer code \cite{WDBEMag} in order to determine the value of $S(\theta_{\gamma})$ \cite{gpcorrel,GdMag,Be2007JPG}, as the nuclei of interest were Coulomb-excited with known reaction geometry. The angular correlations were also measured in several cases to verify the calculations and experimental procedures. In particular, the measured angular correlations can determine if there is any significant offset of the $\gamma$-ray detector from the nominal angle. Such offsets have little impact on deduced relative precession angles, but can strongly affect absolute values, since $S$ is a strong function of $\theta_{\gamma}$ at the detection angles used for the precession measurements.

For short-lived states the precession angle is weakly dependent on the level lifetime, which may be taken into account by expressing
\begin{equation}\label{eq:dth}
\Delta\theta = g~\Phi(\tau),
\end{equation}
where $g$ is the nuclear $g$~factor and $\Phi(\tau)$ is given by:
\begin{equation}\label{eq:phi}
\Phi(\tau) = -\frac{\mu_N}{\hbar} \int_{0}^{T}B_{\rm TF}[v(t)]e^{-t/\tau}dt,
\end{equation}
where $\mu_N$ is the nuclear magneton, $B_{\rm TF}[v(t)]$ is the TF strength at the time-dependent ion velocity $v(t)$, $\tau$ is the mean life of the state of interest, $T$ is the transit time of the nucleus through the ferromagnetic medium, and $t$ is time.

\begin{figure}[]
\centerline{
  \includegraphics[width=\columnwidth]{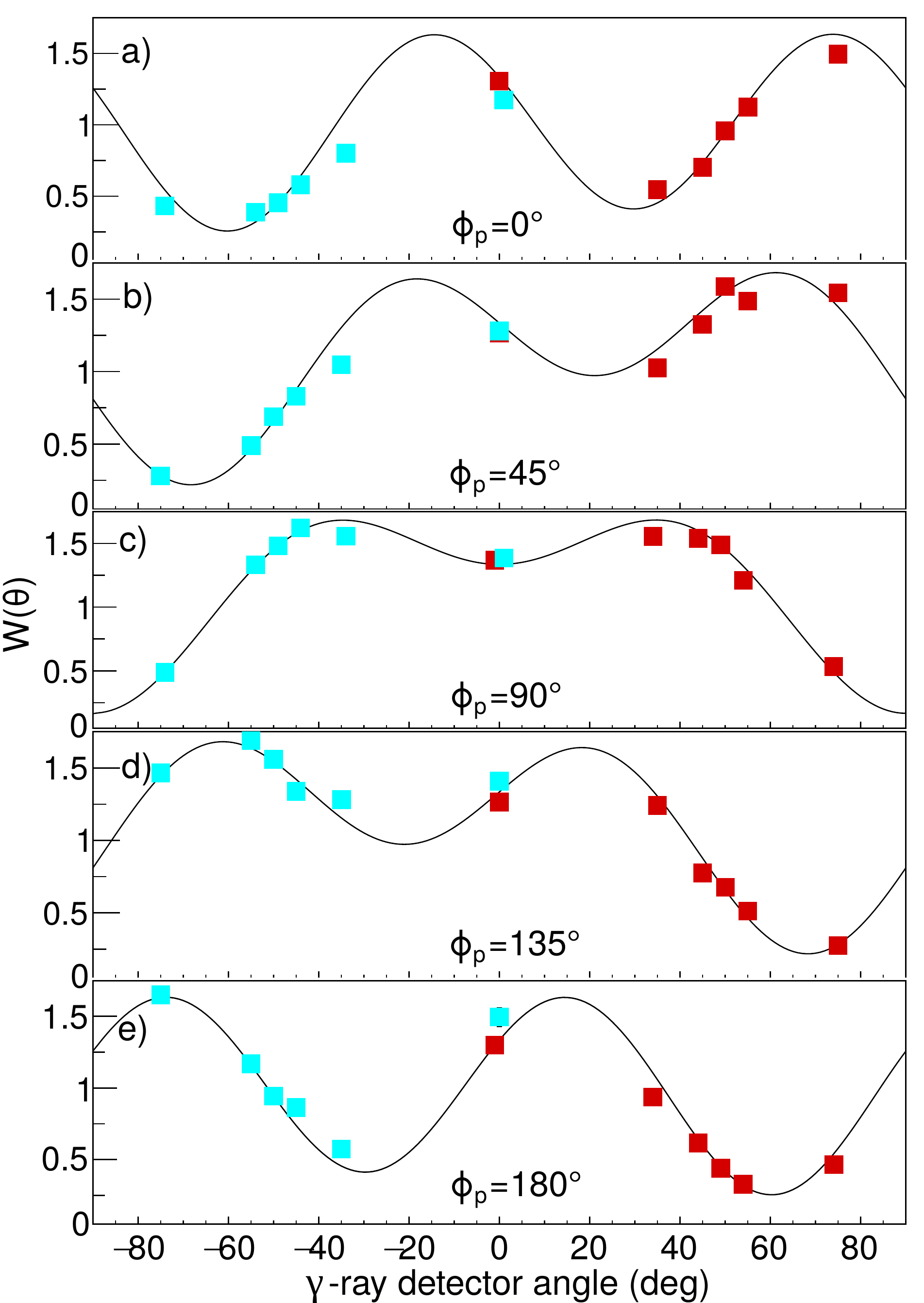}
}
\caption{Angular correlations for the $^{74}$Ge $2^+_1 \rightarrow 0^+_1$ transition from run~3, with measurements taken at varying $\gamma$-ray detector angles for the five particle detector angles. a) Particle detector at $\phi_p=0^{\circ}$. b) Particle detectors at $\phi_p= \pm 45^{\circ}$. c) Particle detectors at $\phi_p= \pm 90^{\circ}$. d) Particle detectors at $\phi_p= \pm 135^{\circ}$. e) Particle detector at $\phi_p=180^{\circ}$. The measured data are normalized to the theoretical angular correlations. Dark red (light blue) points correspond to the $\gamma$-ray detector that moves to positive (negative) angles. Note the reflection symmetry between the correlations for $\phi_p=0$ and $\phi_p=180^{\circ}$, $\phi_p=45^{\circ}$ and $\phi_p=135^{\circ}$.
}
\label{fig:Ge15AC}
\end{figure}

\begin{figure}[]
\centerline{
  \includegraphics[width=\columnwidth]{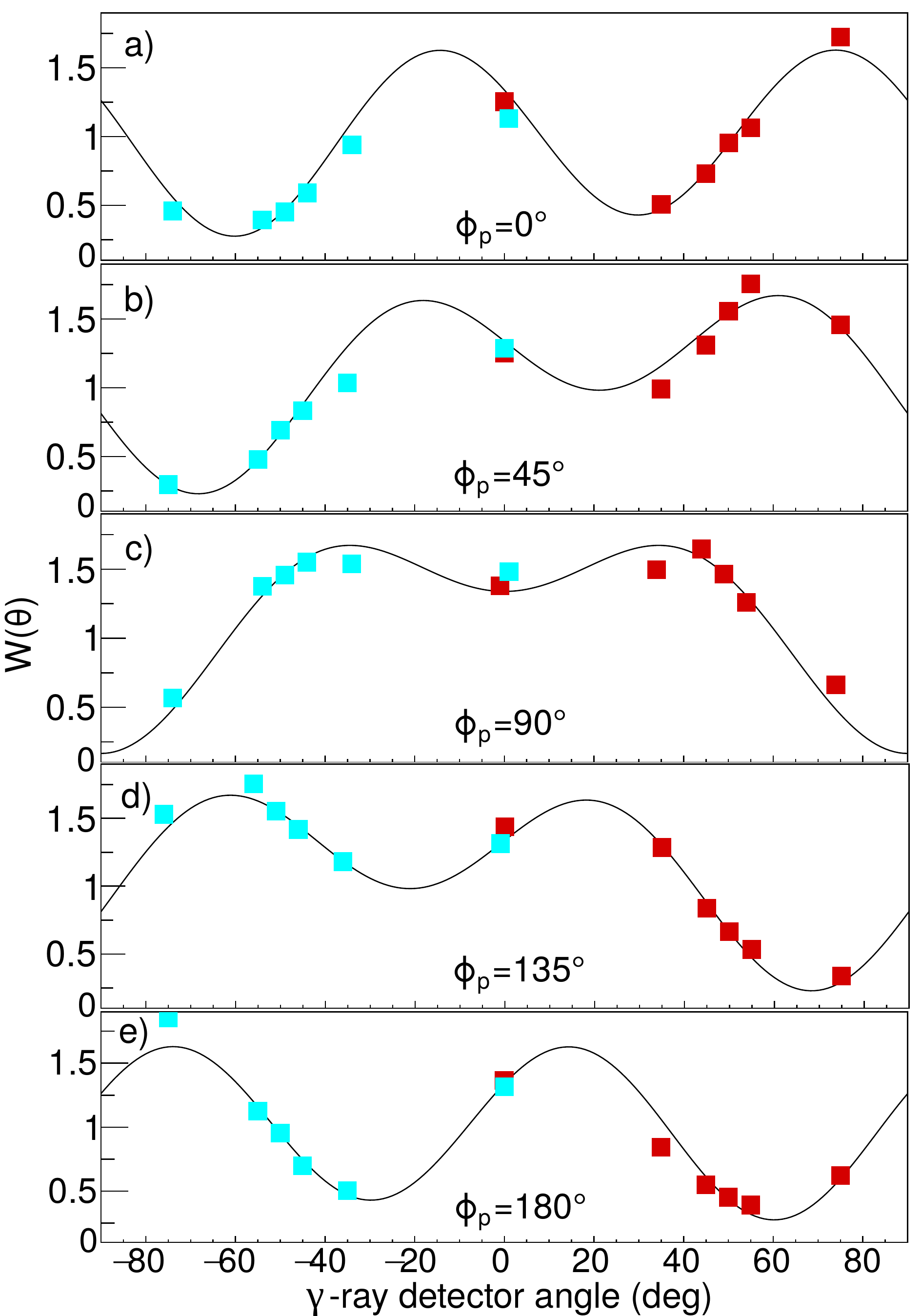}
}
\caption{As for Fig.~\ref{fig:Ge15AC}, for the $^{74}$Se $2^+_1 \rightarrow 0^+_1$ transition in run~3.}
\label{fig:Se15AC}
\end{figure}

The TF strength has been parametrized in terms of the ion velocity and atomic number by fitting an empirical function to data. The most widely used parametrizations have the general form:
\begin{equation}\label{eq:Btf}
B_{\rm TF}[v(t)] = a_{\rm TF} Z^{p_v}(v/v_0)^{p_Z}%e^{-\beta(v/v_0)},
\end{equation}
where $v(t)$ is the time-dependent ion velocity, $Z$ is the atomic number of the ion, $v/v_0$ is the ion velocity in atomic units, $a_{\rm TF}$ is a fitted scale parameter, $p_v$ defines the velocity dependence and $p_Z$ describes the atomic number dependence. Differences in the parameters arise from fitting data sets which cover different atomic-number ranges and span different velocity ranges. However, all parametrizations applicable to ion velocities in the regime considered here have $p_Z \approx 1$. For example, the linear parametrization \cite{EbTFParam} has $p_Z=1$ and the Rutgers parametrization \cite{RutgersTFParam} has $p_Z=1.1$.

The primary objective in the present work is to measure $g$-factor ratios for the Ge and Se isotopes. These are obtained by combining Eq. (\ref{eq:dthexp}) and Eq. (\ref{eq:dth}) to give
\begin{equation}\label{eq:gratio}
\frac{g_x}{g_y} = \frac{\epsilon_x}{\epsilon_y} \frac{S_y}{S_x} \frac{\Phi_y}{\Phi_x} = \frac{\Delta\theta_x}{\Delta\theta_y} \frac{\Phi_y}{\Phi_x},
\end{equation}
where $x$ and $y$ signify the two states being measured. The ratio ${\Phi_y}/{\Phi_x}$ must be evaluated based on a parametrization of the TF strength. For this purpose we have adopted the Rutgers parametrization \cite{RutgersTFParam}, for which $a_{\rm TF}(Fe) = 16.9$~T for fully saturated iron hosts, $a_{\rm TF}(Gd) = 16.6$~T for our gadolinium hosts \cite{ROBINSON1999}, and $p_v=0.45$ and $p_Z=1.1$ for both hosts.

The ratio ${\Phi_y}/{\Phi_x}$  is almost independent of any reasonable choice of parameters. The scale parameter $a_{\rm TF}$ cancels in any ratio, as does the atomic number dependence for ratios within an isotope chain. For nearby atomic numbers, such as for the ratio $\Phi(^{74}{\rm Se})/\Phi(^{74}{\rm Ge})$, the difference between taking $p_Z=1.1$ versus $p_Z=1$ is negligible (0.6\% in this case). $\Phi_y(\tau_y)/\Phi_x(\tau_x)$ is therefore sensitive only to the parameter $p_v$, and then only if one of the lifetimes is short compared to the transit time of the ion through the ferromagnetic foil. In the cases encountered here, the difference in ${\Phi_y}/{\Phi_x}$ that comes about from taking $p_v=1$ versus $p_v=0.45$ is altogether negligible. Furthermore, ${\Phi_y}/{\Phi_x}$ is near unity for most of the cases reported here.

\section{Results} \label{sect:Results}

\begin{figure}[]
\centerline{
  \includegraphics[width=\columnwidth]{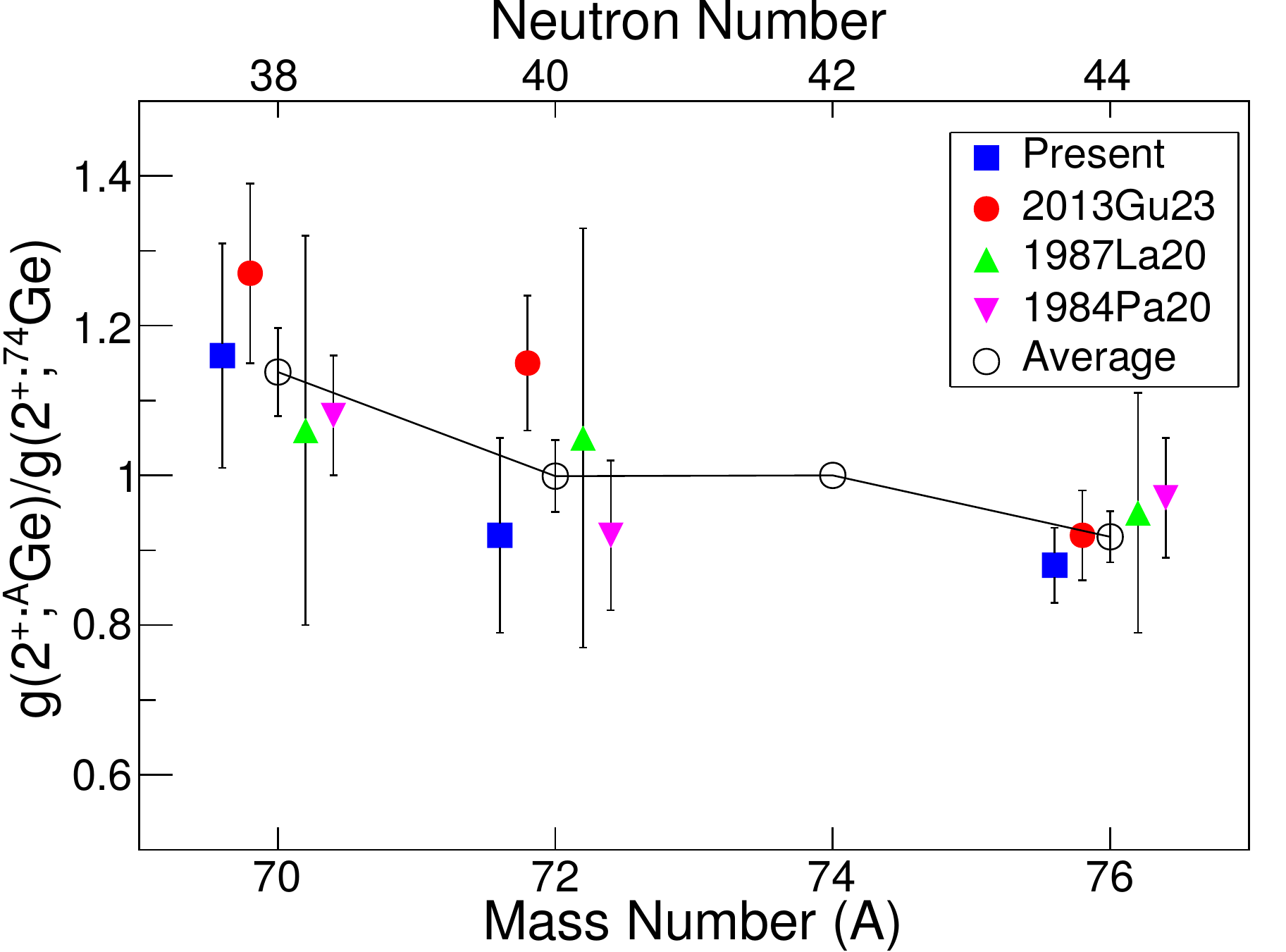}
}
\caption{Ratios of $g(2^+_1)$ in the Ge isotopes relative to $^{74}$Ge, from present and previous work: 1984Pa20~\cite{Pa1984JPG}, 1987La20~\cite{La1987AJP}, and 2013Gu23~\cite{Gu2013PRC}. Values from Ref.~\cite{Gu2013PRC} are taken from their target II measurements only.}
\label{fig:Geratios}
\end{figure}

\begin{figure}[]
\centerline{
  \includegraphics[width=\columnwidth]{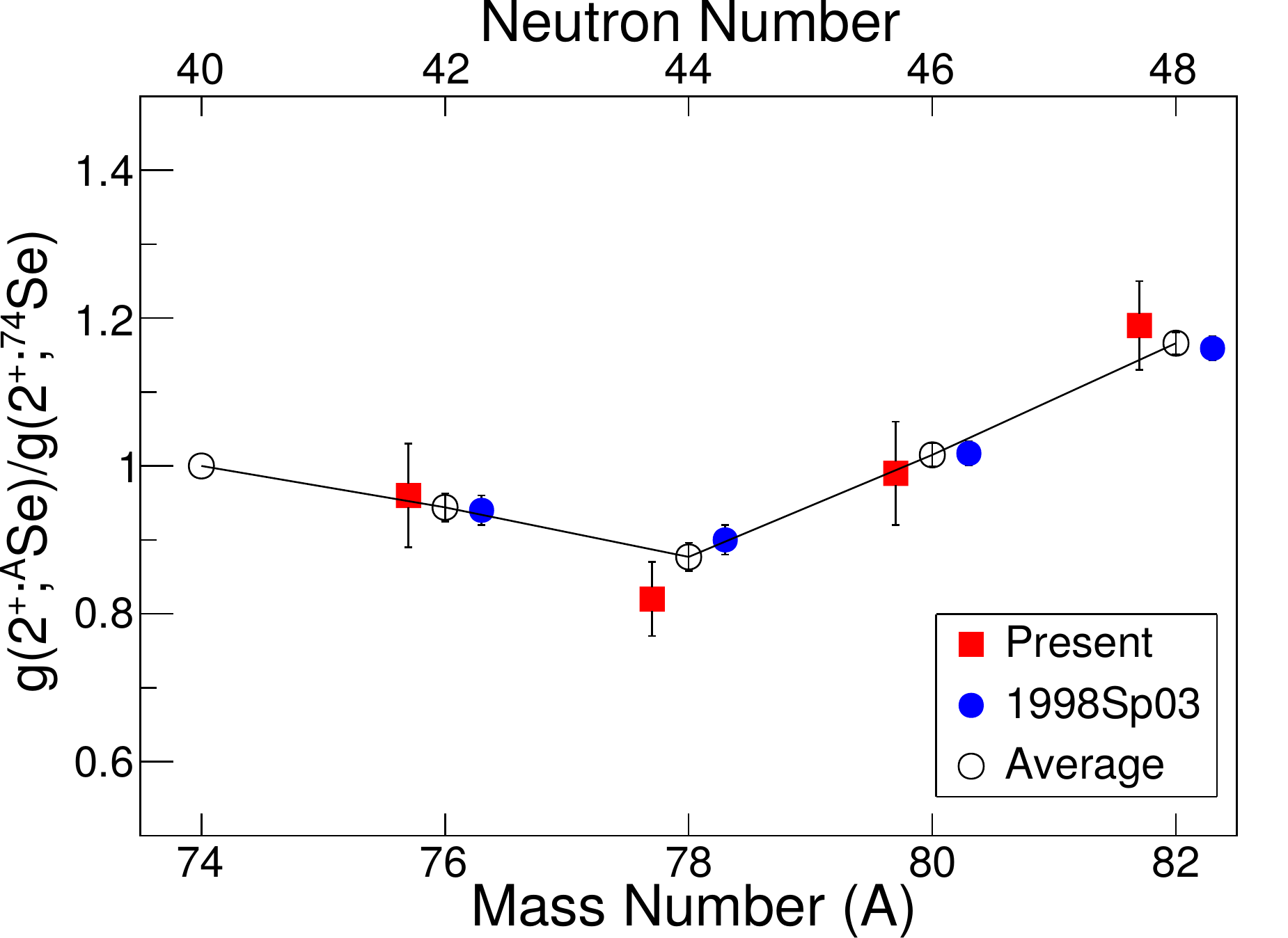}
}
\caption{Ratios of $g(2^+_1)$ in the Se isotopes relative to $^{74}$Se, from present and previous work: 1998Sp03~\cite{Sp1998PRC}.}
\label{fig:Seratios}
\end{figure}

In order to evaluate the $g$-factor ratios, photopeak counts were obtained from background-subtracted integrals of the photopeak region of the $2^+_1 \rightarrow 0^+_1$ transition, after random-coincidence subtraction in particle-$\gamma$ coincidence spectra. An example of a random-subtracted particle-$\gamma$ coincidence spectrum for $^{74}$Ge and $^{74}$Se from run~1 is shown in Fig.~\ref{fig:photopeak}.

Selected measured angular correlations are shown in Figs.~\ref{fig:GeSe18AC}$-$\ref{fig:Se15AC}. Figures \ref{fig:GeSe18AC} and \ref{fig:Ge76AC} compare the angular correlations for the central and outer particle detectors in runs 1 and 2 (see Fig.~\ref{fig:tripledetector}). There is a less pronounced dip at $\theta_{\gamma}=0^{\circ}$ for the outer detectors, but the slope at $\theta_{\gamma}=\pm65^{\circ}$ is very similar for both the central and outer detectors. These figures also show that the calculated angular correlations agree very well with experiment and that the angular correlations are almost identical between the isotopes measured. Thus, the ratio $S_y/S_x$ in Eq.~(\ref{eq:gratio}) is very near unity in all cases considered here.

Figures \ref{fig:Ge15AC} and \ref{fig:Se15AC} show angular correlations for the five particle detectors at $\phi_p = 0^{\circ}$, $45^{\circ}$, $90^{\circ}$, $135^{\circ}$, and $180^{\circ}$ in run 3 (see Fig.~\ref{fig:heliotrope}). The similarity of the angular correlations for  $\phi_p = 90^{\circ}$ to those for the outer detectors in Fig.~\ref{fig:GeSe18AC} and Fig.~\ref{fig:Ge76AC} can be noted. Also evident in Fig.~\ref{fig:Ge15AC} and Fig.~\ref{fig:Se15AC} is that for $\phi_p = 0^{\circ}$, $45^{\circ}$, $135^{\circ}$, and $180^{\circ}$, the angular correlation has a pronounced slope at $\theta_{\gamma} = \pm 45^{\circ}$ and $\theta_{\gamma} = \pm 90^{\circ}$, in some cases of similar magnitude to that at $\theta_{\gamma} = \pm 65^{\circ}$ for the conventionally used $\phi_p = 90^{\circ}$. These angles of maximum slope determined the placement of $\gamma$-ray detectors indicated for runs 3-5 in Table~\ref{tab:runs}.

Precession angles were evaluated as per Eq.~(\ref{eq:dthexp}), with $S(\theta_{\gamma})$ calculated from angular correlation theory \cite{gpcorrel,ggcorrel,GdMag}. The measured precession angles are listed in Table~\ref{tab:kinematics}. In order to check that systematic errors were minimal, cross-ratios~\cite{Be2007JPG} were calculated for all particle-$\gamma$ pairs. The cross-ratios showed a normal distribution around unity, indicating no evidence of systematic error.

Indirect population of the 2$^+_1$ state by feeding from the $4^+_1$ states was determined from the $^{74,76}$Ge and $^{74}$Se spectra as these nuclides have the lowest $4^+_1$-state excitation energies. Population of the $4^+_1$ state was measured to be no greater than $\sim 2\%$ of the $2^+_1$ state population in the strongest case, and therefore the effect of feeding on the measurement of the $2^+_1$-state precession angle could be neglected in all cases.

%Table~\ref{tab:gratio} is up to date 15/8/19

% ANU relative Ge g factors:
% NO.   NAME        VALUE          ERROR      NEGATIVE      POSITIVE
%   1    g Ge70      1.1629       0.15248      -0.15248       0.15248
%   2    g Ge72     0.91757       0.12869      -0.12869       0.12869
%   3    g Ge76     0.87948       0.53839E-01  -0.53839E-01   0.53839E-01

%ANU g factor ratios relative to 74Ge:
%  NO.   NAME        VALUE          ERROR      NEGATIVE      POSITIVE
%   1    g Ge70      1.1629       0.15248      -0.15249       0.15248
%   2    g Ge72     0.91757       0.12869      -0.12868       0.12869
%   3    g Ge76     0.87948       0.53839E-01  -0.53833E-01   0.53844E-01
%   4    g Se74      1.3378       0.71269E-01  -0.71285E-01   0.71254E-01
%   5    g Se76      1.3046       0.10656      -0.10658       0.10655
%   6    g Se78      1.1011       0.69646E-01  -0.69082E-01   0.70192E-01
%   7    g Se80      1.3562       0.97413E-01  -0.96926E-01   0.97887E-01
%   8    g Se82      1.6235       0.95060E-01  -0.93801E-01   0.96336E-01

% ANU relative Se g factors:
% EXT PARAMETER                  PARABOLIC         MINOS ERRORS
%  NO.   NAME        VALUE          ERROR      NEGATIVE      POSITIVE
%   1    g Se74      1.0000       constant
%   2    g Se76     0.96338       0.69340E-01  -0.69345E-01   0.69335E-01
%   3    g Se78     0.81753       0.47284E-01  -0.46979E-01   0.47577E-01
%   4    g Se80     0.99131       0.65795E-01  -0.65556E-01   0.66029E-01
%   5    g Se82      1.1932       0.63694E-01  -0.62900E-01   0.64489E-01

The $g$-factor ratios determined from the measured precession angles and $\Phi(\tau)$ values in Table~\ref{tab:kinematics} are listed in Table~\ref{tab:gratio}, and shown in Fig.~\ref{fig:Geratios} (Ge isotopes) and Fig.~\ref{fig:Seratios} (Se isotopes). The Ge isotopes are referenced to $^{74}$Ge, the Se isotopes are referenced to $^{74}$Se, and the Ge and Se isotopes are related through the ratio $g(^{74}{\rm Se})/g(^{74} {\rm Ge})$. The value of this ratio from the cocktail beam data in runs 1 and 3 alone is $g(^{74}{\rm Se})/g(^{74}{\rm Ge})=1.325(77)$. With the additional data collected for $^{74}$Ge alone from run 1 the ratio becomes $g(^{74}{\rm Se})/g(^{74}{\rm Ge})=1.338(71)$, which is the value given in Table~\ref{tab:gratio}. Runs~4 and~5 did not include either of the reference isotopes, $^{74}$Ge and $^{74}$Se, but they give independent measures of  $g(^{80}{\rm Se})/g(^{82}{\rm Se})$ and $g(^{78}{\rm Se})/g(^{82}{\rm Se})$, respectively. These data were included in the evaluation of the $g$-factor ratios for the Se isotopes relative to $^{74}$Se by performing a chi-squared fit to the complete data set. Such a procedure gives the correct average values with the correct uncertainties in a straight-forward way.

\begin{figure}[]
\centerline{
  \includegraphics[width=\columnwidth]{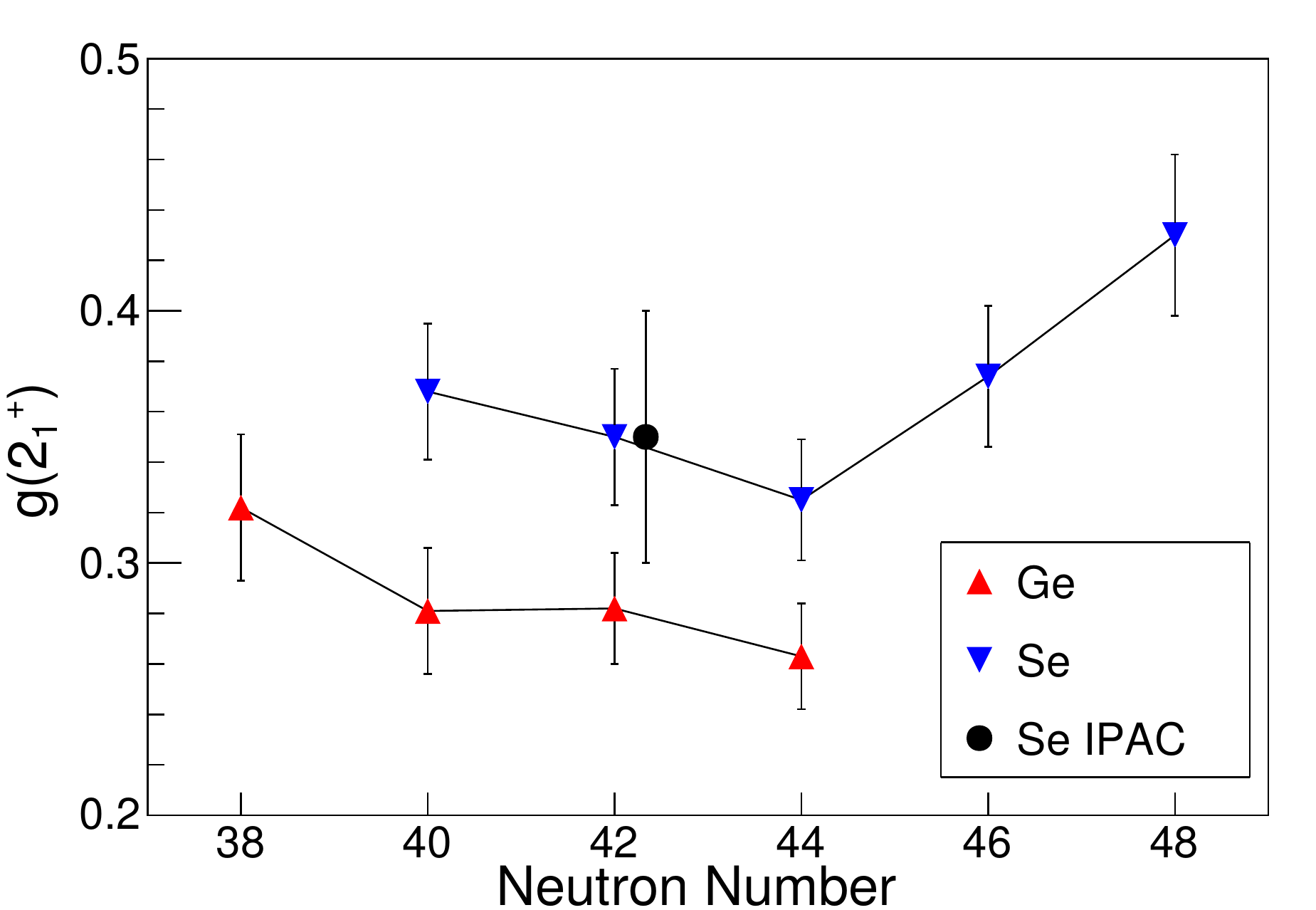}
}
\caption{Comparison of the adopted Ge and Se $g$-factor values  (Table~\ref{tab:gfactors}). There is a shift in magnitude between the two isotope chains but the trend with neutron number appears to be similar. The independently measured IPAC result for $^{76}$Se \cite{Mu1967CJP} is shown for reference (slightly displaced from $N=42$ for clarity).}
\label{fig:GeSeMeasComp}
\end{figure}

The Rutgers parametrization was used to set the absolute scale of the relative $g$-factor measurements. Rather than normalize to a particular reference $g$~factor, a global fit was performed that included the present precession data in Table~\ref{tab:kinematics} and previous $g$-factor ratios from Table~\ref{tab:gratio}. A 10\% uncertainty was assigned to the TF strengths for the iron and gadolinium hosts. The reduced chi-squared value from this fit, $\chi^2_{\nu} = 1.06$, shows the internal consistency of the present measurements and their consistency with the previous $g$-factor ratios. The resultant `absolute' $g$~ factors are shown in Table~\ref{tab:gfactors} and in Fig.~\ref{fig:GeSeMeasComp}.

In the process of fitting these data and testing the sensitivity of the relative $g$~factors to alternative parametrizations of the TF strength, it was observed that assuming a linear velocity dependence for the TF strength gives almost identical $g$-factor ratios, but has an increased chi-squared value, $\chi^2_{\nu}= 2.0$. Thus, it is clear that the present $g$-factor ratios are not sensitive to the velocity dependence of the transient field.

\begin{table}[]
\centering
\caption{$g(2^+_1)$ ratios from the present and previous work \cite{Pa1984JPG,La1987AJP,Gu2013PRC,Sp1998PRC}.}
\label{tab:gratio}
\begin{tabularx}{\hsize}{lllllr}
\hline
\hline
Ratio & Present & \cite{Pa1984JPG} & \cite{La1987AJP} & \cite{Gu2013PRC,Sp1998PRC} & Average \\
\hline
$\mathrm{^{70}Ge}/\mathrm{^{74}Ge}$ & 1.16(15) & 1.08(8) & 1.06(26) & 1.27(12) & 1.138(59)\\
$\mathrm{^{72}Ge}/\mathrm{^{74}Ge}$ & 0.92(13) & 0.92(7) & 1.05(14) & 1.15(9) & 0.999(48)\\
$\mathrm{^{76}Ge}/\mathrm{^{74}Ge}$ & 0.88(5) & 0.97(7) & 0.95(13) & 0.92(6) & 0.918(34)\\
$\mathrm{^{74}Se}/\mathrm{^{74}Ge}$ & 1.34(7) & & & 1.22(8) & 1.288(53)\\

$\mathrm{^{76}Se}/\mathrm{^{74}Se}$ & 0.96(7) & & & 0.942(20) & 0.944(19)\\
$\mathrm{^{78}Se}/\mathrm{^{74}Se}$ & 0.82(5) & & & 0.898(22) & 0.877(19)\\
$\mathrm{^{80}Se}/\mathrm{^{74}Se}$ & 0.99(7) & & & 1.017(16) & 1.015(16)\\
$\mathrm{^{82}Se}/\mathrm{^{74}Se}$ & 1.19(6) & & & 1.159(16) & 1.166(15)\\
\hline
\hline
\end{tabularx}
\end{table}

%ANU+previous g factor ratios relative to 74Ge:
%  NO.   NAME        VALUE          ERROR      NEGATIVE      POSITIVE
%   1    g Ge70      1.1381       0.59392E-01  -0.59399E-01   0.59385E-01
%   2    g Ge72     0.99947       0.47730E-01  -0.47730E-01   0.47730E-01
%   3    g Ge76     0.91774       0.33595E-01  -0.33591E-01   0.33599E-01
%   4    g Se74      1.2857       0.53215E-01  -0.53230E-01   0.53201E-01

%ANU+previous g factor ratios relative to 74Se:
%  NO.   NAME        VALUE          ERROR      NEGATIVE      POSITIVE
%   1    g Se74      1.0000       constant
%   2    g Se76     0.94363       0.19217E-01  -0.19201E-01   0.19233E-01
%   3    g Se78     0.87706       0.19371E-01  -0.19337E-01   0.19405E-01
%   4    g Se80      1.0149       0.15505E-01  -0.15519E-01   0.15491E-01
%   5    g Se82      1.1661       0.15217E-01  -0.15215E-01   0.15219E-01

\section{Discussion} \label{sect:Discussion}

\begin{table}[]
  \begin{center}
    \caption{Measured $g$~factors in Ge and Se using the TF technique. Measurements are ordered chronologically from left to right. Although not explicitly listed, the signs of all $g$~factors in this table are positive.}
    \label{tab:gfactors}
    \begin{tabular}{ccccc}
      \hline
      \hline
      Nuclide &\cite{Pa1984JPG} & \cite{La1987AJP} & \cite{Sp1998PRC,Gu2013PRC} & Present \footnotemark[1] \\
      \hline
      $^{70}$Ge & 0.468(26) & 0.370(89) & 0.44(4)   & 0.322(29) \\
      $^{72}$Ge & 0.399(33) & 0.367(44) & 0.44(2)   & 0.281(25) \\
      $^{74}$Ge & 0.433(20) & 0.350(22) & 0.35(1)   & 0.282(22) \\
      $^{76}$Ge & 0.419(23) & 0.334(39) & 0.32(1)   & 0.263(21) \\
      $^{74}$Se &           &           & 0.428(27) & 0.368(27) \\
      $^{76}$Se\footnotemark[2] & &     & 0.403(23) & 0.350(27) \\
      $^{78}$Se &   &                   & 0.384(25) & 0.325(24) \\
      $^{80}$Se &   &                   & 0.435(27) & 0.374(28) \\
      $^{82}$Se &   &                   & 0.496(29) & 0.430(32) \\
      \hline
      \hline
    \end{tabular}
  \end{center}
  \begin{flushleft}
  \footnotetext[1]{These results are from a global fit to the present data in Table~\ref{tab:kinematics} together with previous $g$-factor ratios from Table~\ref{tab:gratio}. The uncertainties include $\pm 10\%$ uncertainty on the TF parametrizations for the iron and gadolinium hosts.}
  \footnotetext[2]{The only independently determined $g$~factor, by an IPAC measurement \cite{Mu1967CJP}, gives $g=+0.35(5)$ for $^{76}$Se (see text).}
  \end{flushleft}
\end{table}

%  NO.   NAME        VALUE          ERROR      NEGATIVE      POSITIVE
%   1    g Ge70     0.32208       0.29377E-01  -0.27726E-01   0.31157E-01
%   2    g Ge72     0.28070       0.25098E-01  -0.23665E-01   0.26640E-01
%   3    g Ge74     0.28198       0.21831E-01  -0.20428E-01   0.23328E-01
%   4    g Ge76     0.26327       0.21405E-01  -0.20087E-01   0.22817E-01
%   5    g Se74     0.36805       0.27403E-01  -0.25604E-01   0.29333E-01
%   6    g Se76     0.35008       0.26820E-01  -0.25095E-01   0.28667E-01
%   7    g Se78     0.32486       0.24353E-01  -0.22768E-01   0.26052E-01
%   8    g Se80     0.37433       0.28313E-01  -0.26472E-01   0.30284E-01
%   9    g Se82     0.42994       0.32177E-01  -0.30072E-01   0.34432E-01
%  10    f_Fe       0.96125       0.71280E-01  -0.70722E-01   0.71496E-01
%  11    f_Gd        1.0360       0.75598E-01  -0.75171E-01   0.75717E-01

\subsection{$g$-factor measurements} \label{ssect:gfactor}

The $g$-factor values measured previously by the TF technique are shown in Table~\ref{tab:gfactors}. For the Ge isotopes, there have been measurements by the Oxford \cite{Pa1984JPG},  Melbourne \cite{La1987AJP},  and Rutgers \cite{Gu2013PRC} groups based on the Rutgers parametrization. Scrutiny of the $g(2^+_1;\mathrm{^{74}Ge})$ values, having the smallest relative uncertainty, indicates that the Oxford measurement generally gave a larger magnitude for the $^{70-76}$Ge $g$~factors than obtained in the Melbourne and Rutgers measurements. However, as shown in Fig.~\ref{fig:Geratios}, the relative $g$~factors of the previous and present measurements are in overall good agreement. In making this comparison, the data from the Rutgers group \cite{Gu2013PRC} were restricted to those taken using the same target (their target II with gadolinium as the ferromagnetic host). Fig.~\ref{fig:Seratios} shows overall good agreement between the present and previous \cite{Sp1998PRC} $g$-factor ratios for the Se isotopes.

Concerning the ratio of the $g$~factors between the Se and Ge isotopes, the most precise measurements in previous work for $^{74}$Se and $^{74}$Ge are $g$~=~$+0.428(27)$ \cite{Sp1998PRC} and $g$~=~$+0.35(1)$ \cite{Gu2013PRC}, respectively, giving $g(2^+_1;\mathrm{^{74}Se})/g(2^+_1;\mathrm{^{74}Ge})$~=~$1.22(8)$. The measurements were both made using gadolinium as the ferromagnetic host, sampling the TF in overlapping but different velocity ranges. The present simultaneous measurement ensures that the ions sample the TF over effectively the same velocity range, and gives $g(2^+_1;\mathrm{^{74}Se})/g(2^+_1;\mathrm{^{74}Ge})$~=~$1.34(7)$, which agrees with the previous work (the error bars overlap).

From these results two conclusions may be drawn. First, the TF parametrization assumed in the previous work appears to describe the velocity dependence of Ge and Se ions traversing gadolinium hosts within the precision of the data. Second, the TF strength must change smoothly between $Z$~=~$32$ and $Z$~=~$34$ for both iron and gadolinium hosts. The agreement of ratios between present and previous work gives confidence in adopting the weighted average of present and previous $g$-factor \textit{ratios}. The remaining challenge then is to determine the \textit{absolute} scale.

The IMPAC data on the Ge and Se isotopes \cite{He1969NPA} cannot determine the absolute $g$~factors, mainly because the static hyperfine field takes several picoseconds to reach its full strength following implantation \cite{An1995HI,St1999PRL}. This equilibration time ($\approx 5$~ps) is a significant fraction of the 2$^+_1$-level lifetime in all of the isotopes studied. All that the IMPAC data can provide is a lower estimate on the absolute $g$~factors by making the extreme assumptions that all implanted ions reside on full-field sites and that the static hyperfine field reaches its full strength promptly after implantation. Under these conditions the absolute $g$~factors would be about 20\% smaller than those in the last column of Table~\ref{tab:gfactors}. More realistic assumptions about the effective hyperfine fields at the implanted ions readily accommodate both the present and previous $g$~factor values. (This discussion of the IMPAC data has excluded the $^{82}$Se measurement, which is incompatible with the $g$-factor ratios reported here and in Ref.~\cite{Sp1998PRC}.)

To our knowledge the only independently measured $g$~factor that can serve to set the magnitude of the Ge and Se $g$~factors is an integral $\gamma \gamma$ perturbed-angular-correlation~(IPAC) measurement on $^{76}$Se \cite{Mu1967CJP}, populated by the radioactive decay of $^{76}$As. With a more recently measured static-hyperfine-field strength of $B_{\rm hf}$~=~$67.9(10)$~T~\cite{St2001HI} and lifetime of $\tau$~=~$17.75(29)$~ps~\cite{RamanBE2}, this measurement gives $g(2^+_1;\mathrm{^{76}Se})$~=~$+0.35(5)$, i.e.,  with 14\% uncertainty.  The present `absolute' $g$~factor for $^{76}$Se, $g=+0.350(27)$, happens to agree perfectly with this value. The IPAC $g$~factor is smaller than, but nevertheless consistent with, $g(2^+_1;\mathrm{^{76}Se})$~=~$+0.403(23)$ as reported in Ref.~\cite{Sp1998PRC}.

The present `absolute' $g$-factor values are about 80\% of those reported previously. It is not certain why, given that the present and previous measurements are all based more or less directly on the Rutgers parametrization. There are, however, two important differences between the present and previous work: (i) the ion velocities in the present work are higher than those used to determine the Rutgers parametrization, and (ii) the present work primarily used iron rather than gadolinium hosts. The more recent statistically precise measurements~\cite{Sp1998PRC,Gu2013PRC} used gadolinium hosts whereas iron hosts were chosen for the present measurements to help avoid the possible beam heating effects that have been noted by Benczer-Koller and Kumbartzki \cite{Be2007JPG}. Beam heating is more likely to affect gadolinium hosts with a Curie temperature of 293~K than iron hosts with Curie temperature 1043~K.

In summary: the present and previous relative $g$-factor measurements are robust, but the absolute scale remains uncertain on the level of $\pm 20\%$. Nevertheless the `absolute' $g$~factors adopted here agree well with the only independent $g$-factor measurement on $^{76}$Se, and so we proceed to compare these values with theory.

\begin{figure}[]
\centerline{
  \includegraphics[width=\columnwidth]{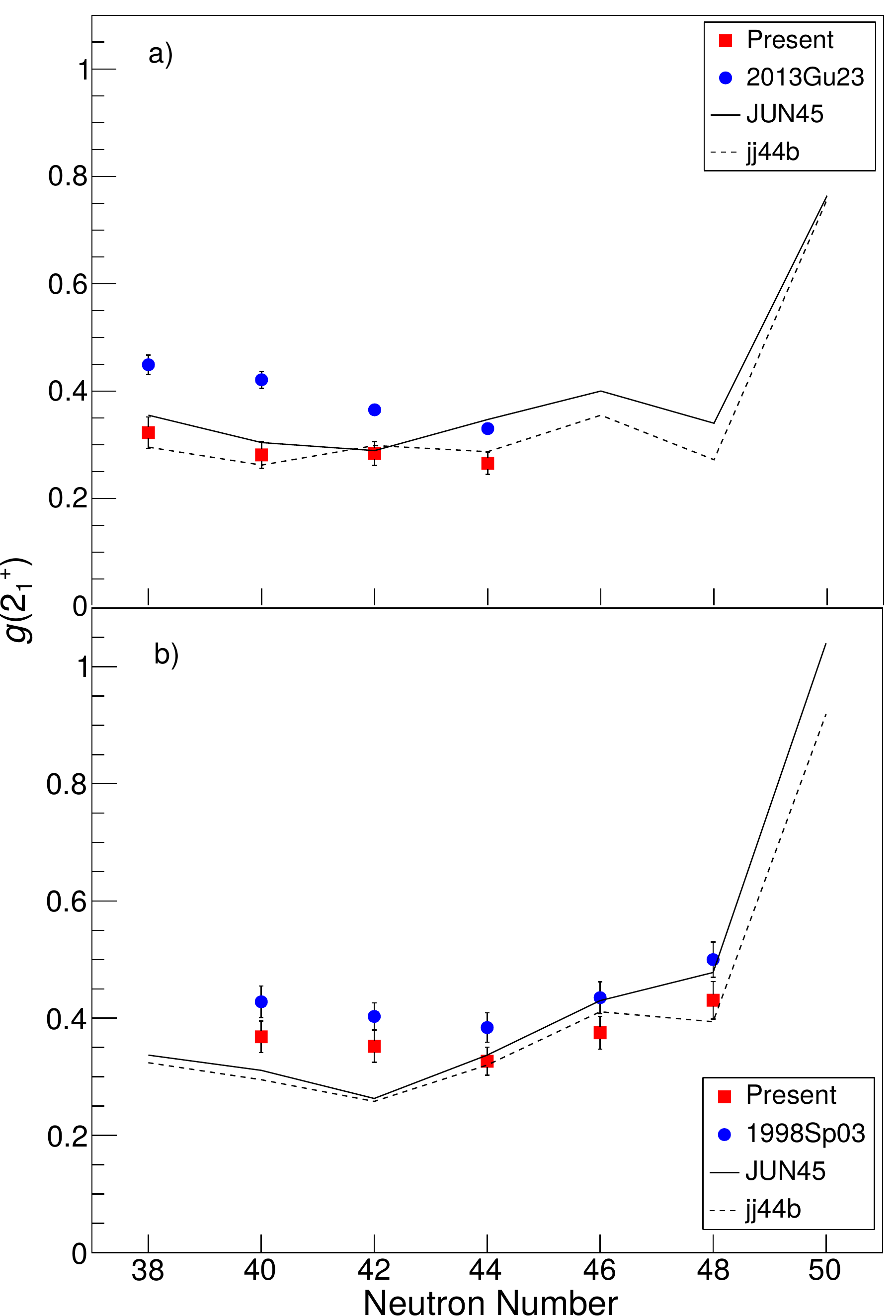}
}
\caption{Measured 2$^+_1$-level $g$~factors for a) Ge and b) Se isotopes from the present work (red squares) and the literature (blue circles). Previous work is designated by the Nuclear Science Reference code, namely 2013Gu23~\cite{Gu2013PRC} and 1998Sp03~\cite{Sp1998PRC}. Shell-model calculations using the JUN45 and jj44b interactions are shown as solid lines and  dashed lines, respectively. Experimental $g$~factors from Ref.~\cite{Gu2013PRC} shown in panel a) are their reported weighted-averages that include the previous measurements \cite{Pa1984JPG,La1987AJP} along with their own data. (See also Table~\ref{tab:gfactors}.)}
\label{fig:GeSeexpvscalcg}
\end{figure}

\begin{table*}[t]
  \begin{center}
    \caption{Component breakdown of the Ge and Se $2^+_1$-state magnetic-moments ($\mu$) for the JUN45 interaction \cite{Ho2009PRC}.}
    \label{tab:JUN45}
    \begin{tabularx}{\textwidth}{XXXXXXXXXXX}
      \hline
      \hline
      $Z$ & $A$ & $N$ & $\langle L_{p} \rangle$ & $\langle S_{p} \rangle$ & $\langle L_{n} \rangle$ & $\langle S_{n} \rangle$ & $\mu_p$ & $\mu_n$ & $\mu$ & $g(2^+)$ \\
      \hline
      32 & 70 & 38 & 0.772 &  0.0044 & 1.192 & 0.0297  & 0.790 & -0.079 & 0.710 & 0.355 \\
      32 & 72 & 40 & 0.760 &  0.0023 & 1.176 & 0.0605  & 0.769 & -0.162 & 0.607 & 0.304 \\
      32 & 74 & 42 & 0.770 &  0.0032 & 1.150 & 0.0764  & 0.782 & -0.205 & 0.578 & 0.289 \\
      32 & 76 & 44 & 0.897 &  0.0039 & 1.018 & 0.0813  & 0.912 & -0.218 & 0.694 & 0.347 \\
      32 & 78 & 46 & 1.017 &  0.0027 & 0.895 & 0.0852  & 1.028 & -0.228 & 0.800 & 0.400 \\
      32 & 80 & 48 & 1.002 & -0.0120 & 0.907 & 0.1033  & 0.955 & -0.277 & 0.679 & 0.339 \\
      32 & 82 & 50 & 2.162 & -0.1622 & 0.000 & 0.0000  & 1.528 &  0.000 & 1.528 & 0.764 \\
      \\
      34 & 72 & 38 & 0.847 & -0.0123 & 1.110 & 0.0469  & 0.799 & -0.126 & 0.673 & 0.337\\
      34 & 74 & 40 & 0.814 & -0.0104 & 1.132 & 0.0568  & 0.773 & -0.152 & 0.621 & 0.310 \\
      34 & 76 & 42 & 0.745 & -0.0025 & 1.174 & 0.0787  & 0.735 & -0.211 & 0.525 & 0.262 \\
      34 & 78 & 44 & 0.866 &  0.0064 & 1.044 & 0.0814  & 0.891 & -0.218 & 0.674 & 0.337 \\
      34 & 80 & 46 & 1.044 &  0.0090 & 0.865 & 0.0822  & 1.079 & -0.220 & 0.859 & 0.429 \\
      34 & 82 & 48 & 1.163 &  0.0064 & 0.744 & 0.0869  & 1.188 & -0.233 & 0.955 & 0.478 \\
      34 & 84 & 50 & 1.972 &  0.0279 & 0.000 & 0.0000  & 2.081 &  0.000 & 2.081 & 1.041 \\
      \hline
      \hline
    \end{tabularx}
  \end{center}
\end{table*}

\subsection{Shell-model calculations} \label{ssect:shellmodel}

Shell-model calculations were performed using the shell-model program NuShellX \cite{NuShellX} with the JUN45 \cite{Ho2009PRC} and jj44b \cite{JJ44BUnpublished} interactions. Previous studies with these interactions include a comparison between calculated level energies and $E2$ observables for $^{78-82}$Se by Srivastava and Ermamatov \cite{Sr2013PS}, as well as calculations of level energies and $E2$ transtions for $^{76}$Ge \cite{Mu2017PRC} and $^{76}$Se \cite{Mu2019PRC} by Mukhopadhyay \textit{et al.} The appendix of Ref.~\cite{Mu2017PRC} gives a detailed description of the model space and two-body matrix elements. In brief, both interactions are based on a renormalized Bonn-C potential with the same assumed mass dependence. The differences between the interactions arise from fitting different sets of binding-energy and excitation-energy data. The jj44b interaction fits data for $Z=28-30$ ($N=28-50$) and also for $N=48-50$ ($Z=28-50$), whereas JUN45 fits nuclei in the ranges $Z=30-32$ ($N=32-50$) and $N=46-50$ ($Z=31-46$). In relation to the present results, JUN45 includes all Ge isotopes and $^{80,82}$Se in its fits, whereas jj44b includes only $^{82}$Se.

The effective $M1$ operator $g_s^{\rm eff}$~=~$0.7 g_s^{\rm free}$ and $g_l^{\rm eff}$~=~$g_l^{\rm free}$ for both protons and neutrons was taken from Ref.~\cite{Ho2009PRC}.  A breakdown of the spin contributions to the $g$~factors is given for the JUN45 interaction in Table~\ref{tab:JUN45}. The theoretical $g$~factors are compared with experiment in Fig.~\ref{fig:GeSeexpvscalcg}. As the absolute values of the $g$~factors are not well-determined experimentally, we focus on the trends. Both interactions predict a marked increase in the $g$~factor at $N=50$, the neutron shell closure. Among the observed $g$~factors, the smallest value occurs at $N=44$ for both isotope chains, however for the Ge isotopes there are no data beyond $N=44$ to determine if the downward trend observed from $^{70}$Ge to $^{76}$Ge continues, or whether the $g$~factors begin to increase as both calculations predict. The qualitative v-shaped trend observed in the mass dependence of the Se isotopes is evident in both calculations, which agree with each other from $^{74}$Se to $^{80}$Se, but place the smallest $g$~factor at $N=42$, rather than at $N=44$ as observed experimentally. This observation of the smallest $g$~factor being at $N=44$, which is the mid point of the neutron $p_{1/2}g_{9/2}$ ``shell",  has been interpreted in terms of the interacting boson model as evidence for a firm (sub)shell closure at $N=38$ along with the well-established shell closure at $N=50$ \cite{Sp1998PRC}.

The $g$~factor can be broken down into its contributions from the orbital ($L$) and spin ($S$) angular momenta of the protons and neutrons. Specifically, $g(2^+) = \mu(2^+)/2$, where the magnetic moment is separated into a proton and a neutron component: $\mu = \mu_p + \mu_n$ with $\mu_p = g_l(p) \langle L_p \rangle + g_s(p) \langle S_p \rangle$, and likewise for $\mu_n$. Inserting the effective nucleon $g$-factor values gives
\begin{equation}
g(2^+) =  (\langle L_p \rangle + 3.91 \langle S_p \rangle - 2.68 \langle S_n \rangle)/2,
\end{equation}
or, noting from Table~\ref{tab:JUN45} that $3.91 \langle S_p \rangle \approx 0$ and $2.68 \langle S_n \rangle \approx 0.2$,
\begin{equation}
g(2^+) \approx  (\langle L_p \rangle -0.2) /2.
\end{equation}
Similar behavior can be anticipated for the jj44b interaction. Thus, the trends shown in Fig.~\ref{fig:GeSeexpvscalcg} for the theoretical $g$~factors reflect primarily the proton orbital contribution to the angular momentum. Small variations in $\langle L_p \rangle$ directly impact on the calculated $g$~factor. Experimentally, the variation in the $g$~factors from the average value along each chain of stable isotopes is on the order of $\sim \pm 0.05$, which corresponds to changes in $\langle L_p \rangle$ on the order of $\sim \pm 0.1$. This estimate can be made despite uncertainties in the absolute magnitude of the $g$~factors.

Neutron scattering studies of $^{76}$Ge \cite{Mu2017PRC} and $^{76}$Se \cite{Mu2019PRC} have allowed the band structures of these isobars to be observed, and low-energy shape coexistence to be identified in $^{76}$Se. Overall, a good description of the low-lying level structures was obtained by the same shell-model calculations as discussed here. Given the considerable sensitivity of the $g(2^+_1)$ values to the balance of the spin carried by orbital motion of the protons, it is suggested that these shell-model calculations should not be expected to reproduce in detail the mass-dependent variation of the $g$~factors of the stable isotopes. Indeed, the stable Ge and Se isotopes are challenging to interpret due to their shape-transitional nature, together with the occurrence of shape coexistence in many of the isotopes \cite{Ay2016PLB,Lj2008PRL,Pe1990NPA, Heyde-Wood-RevModPhys.83.1467}. Evidently, the $g(2^+_1)$ values will be affected by configuration mixing and by the degree of mixing between shapes.

It is relevant to note that the shell-model calculations give a reasonable description of the $E2$ transition strengths in $^{76}$Ge and $^{76}$Se \cite{Mu2017PRC,Mu2019PRC} using standard effective charges for the region \cite{Ho2009PRC}. A firm statement on whether or not there is additional collectivity in the $2^+_1$ states of these nuclei not captured by the shell-model calculations, as suggested by G\"urdal \textit{et al.} \cite{Gu2013PRC}, requires that the absolute values of the $g$~factors be measured accurately and with some precision.

\section{Conclusion} \label{sect:Conclusion}

Relative $g$~factors have been measured for the first-excited states of the stable, even isotopes of Ge and Se. The new relative $g$~factors agree well with relative $g$~factors taken from literature.

Although there remains a need for reliable, independent calibration of the TF strength for Ge and Se ions, the present confirmation of previously-reported $g$-factor ratios also confirms that any marked difference in free-ion hyperfine interactions observed for Ge and Se ions recoiling in vacuum (see Refs.~\cite{RFLthesis,St2013HFI}) must stem from atomic physics and not from the nuclear $g$~factors. Extensive atomic physics calculations are in progress, with the hope that the free-ion fields can be understood well enough in multi-electron systems, such as Na-like ions, that they can be used to perform accurate, absolute $g$-factor measurements on short-lived nuclear states using the time-dependent RIV technique \cite{Mg24RIV,AES_FIG12}. For example, a precise, accurate and absolute measurement of $g(2^+_1;\mathrm{^{74}Ge})$ would set an absolute scale for the present relative $g$-factor measurements.

Even if an absolute RIV measurement remains elusive for the stable isotopes, the present relative $g$~factors, together with extensive studies of the relevant free-ion hyperfine fields \cite{RFLthesis,St2013HFI}, make it feasible to attempt an analysis of the $g$~factors in the neutron-rich isotopes $^{78,80,82}$Ge based on the vacuum deorientation of angular correlations observed following Coulomb excitation~\cite{Pa2005PRL}. Both interactions predict marked variations in the $g$~factors between $N$~=~$46$ and $N$~=~$50$ that may be detectable even with a low intensity radioactive beam measurement~\cite{136TeRIV}.

Comparisons of the measured relative $g$~factors with large-basis shell-model calculations gives a reasonable degree of agreement and suggest that the $g$-factor variations are strongly correlated with the orbital angular momentum carried by the protons. A detailed description of the more subtle mass dependent variations in the $g$~factors of the first-excited states remains a challenge for these complex transitional nuclei that display mixing of prolate, oblate and spherical structures. \\

\begin{acknowledgments}
The authors are grateful to the academic and technical staff of the Department of Nuclear Physics and the Heavy Ion Accelerator Facility (Australian National University) for their continued assistance and maintenance of the facility. Puvaa Rajan, Asif Ahmed and Vincent Margerin are thanked for assistance with a portion of the data collection.
This research was supported in part by the Australian Research Council grant number DP170101673 and the United States National Science Foundation grant number 1811855. B.P.M., T.J.G., M.S.M.G., A.A., B.J.C. and J.T.H.D. acknowledge the support of the Australian Government Research Training Program. Support for the ANU Heavy Ion Accelerator Facility operations through the Australian National Collaborative Research Infrastructure Strategy (NCRIS) program is acknowledged.
\end{acknowledgments}

\bibliographystyle{apsrev4-1}
\bibliography{GeSeRelative}

\end{document}